\documentclass[prd,byrevtex
,preprint
,showkeys
,showpacs
,amsfonts,amsmath,amssymb
]{revtex4}
\usepackage{amsmath,amsthm,amssymb}
\usepackage{bm}
\usepackage[dvipdfmx]{graphicx}
\usepackage{ascmac}
\usepackage{braket}
\usepackage{latexsym}
\usepackage{cases}
\usepackage{physics}
\usepackage{color,float}

\theoremstyle{definition}

\begin{document}
\title{Squeezing of primordial gravitational waves as quantum discord}
\author{Akira Matsumura}
\email{matsumura.akira@h.mbox.nagoya-u.ac.jp}
\author{Yasusada Nambu}
\email{nambu@gravity.phys.nagoya-u.ac.jp}
\affiliation{Department of Physics, Graduate School of Science, Nagoya 
University, Chikusa, Nagoya 464-8602, Japan}
\date{Jan 8, 2020} 

\begin{abstract}
    We investigate the squeezing of primordial gravitational waves
    (PGWs) in terms of quantum discord. We construct a classical state
    of PGWs without quantum discord and compare it with the
    Bunch-Davies vacuum. Then it is shown that the oscillatory
    behavior of the angular-power spectrum of the cosmic microwave
    background (CMB) fluctuations induced by PGWs can be the signature
    of the quantum discord of PGWs. In addition, we discuss the effect of
    quantum decoherence on the entanglement and the quantum
      discord of PGWs  for  super-horizon
      modes. For the state of PGWs with  decoherence effect, we examine
      the decoherence condition and the correlation
      condition  introduced by C.~Kiefer \textit{et al.} (\emph{Class. Quantum Grav.} \textbf{24} (2007) 1699). We show that the decoherence condition is not
      sufficient for the separability of PGWs and the
      correlation condition implies that the PGWs in the
      matter-dominated era have quantum discord.
\end{abstract}
\maketitle

\tableofcontents
\section{Introduction}
In modern cosmology, the early stage of the universe is described
by inflation models. The theory of inflation predicts primordial quantum
fluctuations as the origin of the structure of our universe and
primordial gravitational waves (PGWs). PGWs can be the evidence of
inflation, and its quantum feature is expected to give the information
of quantum gravity.  It is predicted that PGWs generated in the inflation era have the squeezed distribution \cite{Grischuk1989, Grischuk1990}. If 
their statistical feature is observed then it can support
inflation. The detection of the squeezing effect of PGWs by ground-
and space-based gravitational interferometers was discussed by
  B.~Allen, E.~E.~Flanagan and M.~A.~Papa~\cite{Allen2000}. According to their analysis, the detector with a
very narrow band is required to detect the squeezing effect. The
estimated bandwidth is around the present Hubble parameter, and it is
difficult to detect the squeezed property of PGWs practically. On the
other hand, S.~Bose and L.~P.~Grishchuk~\cite{Bose2002}
  considered the indirect observations of squeezing feature of PGWs
by CMB fluctuations. They showed that the squeezing effect appears as
the oscillatory behavior of the angular-power spectrum of the CMB
temperature fluctuations induced by PGWs. This oscillation caused by PGWs is different from the baryon acoustic oscillation induced mainly by primordial density fluctuations. The contribution of PGWs to the acoustic oscillation is very small.

In order to characterize quantum feature of primordial fluctuations,
the notion of quantum correlations is often applied. In particular,
quantum entanglement of primordial fluctuations in the cosmological
background has been investigated \cite{Nambu2008, Nambu2011,
  Maldacena2012, Kanno2014, Kanno2015, Matsumura2018}. In previous
works \cite{Maldacena2012, Matsumura2018}, it was shown that the
entanglement of primordial fluctuations remains during
inflation. Although quantum entanglement is adopted to characterize
the nonlocal properties of quantum mechanics, it describes only a part
of quantum correlations. Quantum discord is a kind of quantum
correlations \cite{Oliver2001,Henderson2001} and is robust against quantum decoherence. In the cosmological context, quantum discord was
investigated in several works
\cite{Nambu2011,Lim2015,Martin2016,Kanno2016,Hollowood2017}.

In this paper, we examine the squeezed nature of PGWs in terms
  of quantum correlations. In the field of quantum information, it
is known that the squeezing of states is related to quantum
correlations. The oscillatory behavior of PGWs originated from the
squeezing can be the evidence of quantum correlation. In order to
clarify the relation between the oscillatory behavior and quantum
correlations, we introduce a classical state of PGWs under several
assumptions. The meaning of classicality is defined based on the
absence of quantum discord. 
The constructed classical state tells us that the oscillatory
feature of PGWs is associated with quantum discord. We compute the
angular-power spectrum of the CMB temperature fluctuations caused by
PGWs and find that there is no oscillatory behaviors for the
classical state of PGWs unlike the Bunch-Davies vacuum. 
Our analysis provides the meaning of the
oscillatory behavior in terms of quantum correlations. We can
regard it as the signature of quantum discord of PGWs.

Furthermore we investigate how the quantum correlation of PGWs
is affected by the quantum decoherence for  super-horizon modes. Under
the assumption that sub-horizon modes of PGWs does not decohere, the decoherence condition and the correlation condition are computed. The
decoherence condition implies the loss of coherence of the
Bunch-Davies vacuum, and the correlation condition means the
sufficient squeezing of the Wigner function for a considering mode in
the phase space. Through the calculation, we show that the decoherence
condition for the super-horizon modes does not mean the
separability of the decohered state of PGWs. We further find that the
correlation condition leads to the survival of the quantum discord of
PGWs in the matter-dominated era.

This paper is organized as follows. In
Sec. \ref{sec:tensorperturbation}, we review the linear theory of a
tensor perturbation of the Friedmann-Lemaitre-Robertson-Walker (FLRW) metric and the oscillatory feature of the correlation function of
the tensor field. In Sec.~\ref{sec:discord}, we construct a classical
state of PGWs and clarify the connection between the oscillatory
behavior of the angular-power spectrum and the quantum discord of
PGWs. In Sec.~\ref{sec:decoherence}, we evaluate the decoherence and
the correlation conditions for the decohered state of PGWs and discuss
the relation to the quantum correlations of PGWs in the matter
era. The section \ref{sec:summary} is devoted to summary. We use the
natural unit $\hbar=c=1$ through this paper.


\section{Quantum tensor perturbation in inflation, radiation and matter era}
\label{sec:tensorperturbation}
In this section, we demonstrate the oscillatory
behavior of the correlation function of PGWs. We consider a 
tensor perturbation of the spatially flat FLRW metric. The
perturbed metric of the spacetime is
\begin{equation}
ds^{2}=a^{2}(\eta)[-d\eta^{2}+(\delta_{ij}+h_{ij})dx^{i}dx^{j}] , \label{eq:perturbedmetric}
\end{equation}
where $\eta$ is the conformal time and $h_{ij}$ represents
the tensor perturbation with $\partial^{j} h_{ij}=\delta^{ij} h_{ij}=0$
($i,j=1,2,3$). We assume that the universe has instantaneous
transitions at $\eta=\eta_\text{r}$ and
$\eta=\eta_\text{m}$ for its expansion law. The scale factor $a$ is given as
\begin{align}
  a(\eta) = 
\begin{cases}
    -\dfrac{1}{H_\text{inf}\,(\eta-2\eta_\text{r})} & (-\infty< \eta
    \leq \eta_\text{r}) \\
    \dfrac{\eta}{H_\text{inf}\,\eta^{2}_\text{r}} &(\eta_\text{r}< \eta
    \leq \eta_\text{m}) \\
    \dfrac{1}{4} \left(1+\dfrac{\eta}{\eta_\text{m}} \right)^{2} \dfrac{\eta_\text{m}}{H_\text{inf}\,\eta_\text{r}^{2}} & (\eta_\text{m} < \eta )
  \end{cases}. \label{eq:FLRWscalefactor}
\end{align}
Each form of the scale factor represents the expansion law in the
inflation, radiation and matter era. The inflationary universe
  is assumed to be the de Sitter spacetime with the Hubble parameter
$H_\text{inf}$. The perturbed Einstein-Hilbert action up to the second
order of $h_{ij}$ is
\begin{equation}
S=\frac{M^{2}_\text{pl}}{2} \int d^{4}x \sqrt{-g} R \approx \frac{M^{2}_\text{pl}}{8} \int d\eta\, d^{3}x\, a^{2} \left(h^{ij}{}' h'_{ij}- \partial^{k} h^{ij} \partial_{k} h_{ij} \right), \label{eq:perturbedEH}
\end{equation}
where prime denotes the derivative of the conformal time $\eta$ and
$M_\text{pl}$ is the reduced Planck mass $1/\sqrt{8\pi G}$. In the
following, we use the rescaled perturbation and its conjugate momentum
\begin{equation}
 y_{ij}:=ah_{ij},\quad \pi_{ij}:=y'_{ij}-\frac{a'}{a} y_{ij}. \label{eq:ypi}
\end{equation}
Since the background spacetime is invariant under spatial rotations and translations, the tensor perturbation can be decomposed as 
\begin{align}
y_{ij}(\bm{x},\eta)&=\frac{\sqrt{2}}{M_\text{pl}} \int \frac{d^{3} q}{(2\pi)^{3/2}} \sum_{\lambda} y_{\lambda}(\bm{q},\eta) e_{ij}(\hat{q},\lambda) e^{i\bm{q}\cdot \bm{x}} ,\label{eq:modeyq} \\
\pi_{ij}(\bm{x},\eta)&=\frac{M_\text{pl}}{\sqrt{2}} \int \frac{d^{3} q}{(2\pi)^{3/2}} \sum_{\lambda} \pi_{\lambda}(\bm{q},\eta) e_{ij}(\hat{q},\lambda) e^{i\bm{q}\cdot \bm{x}} ,\label{eq:modepiq}
\end{align}
where $\lambda=1,2$ denote the labels of the polarization and the polarization
tensor $e_{ij}(\hat{q},\lambda)$ with $\hat{q}=\bm{q}/|\bm{q}|$ is chosen as
\begin{align}
&\hat{q}^{i} e_{ij}(\hat{q},\lambda)=e^{i}{}_{i}(\hat{q},\lambda)=0,\label{eq:tt} \\
&e^{ij*}(\hat{q},\lambda)e_{ij}(\hat{q},\lambda')=2\delta_{\lambda \lambda'}, \label{eq:orthonormal} \\
&e_{ij}^{*}(\hat{q},\lambda)=e_{ij}(-\hat{q},\lambda). \label{eq:parity}
\end{align}
Eq.~\eqref{eq:tt} corresponds
to the traceless and transverse conditions and Eq.~\eqref{eq:orthonormal} 
is the normalization condition. The
representation of the parity transformation for the polarization
tensor is fixed by Eq.~\eqref{eq:parity}. The reality
  condition of the tensor perturbation with \eqref{eq:parity} implies that the variables $y_{\lambda}$ and $\pi_\lambda$ satisfy
\begin{equation}
 y^{*}_{\lambda}(\bm{q},\eta)=y_{\lambda}(-\bm{q},\eta), \quad \pi^{*}_{\lambda}(\bm{q},\eta)=\pi_\lambda (-\bm{q},\eta). \label{eq:parityypi}
\end{equation}
From the perturbed action \eqref{eq:perturbedEH}, the mode equation is
\begin{equation}
y''_{\lambda}(\bm{q},\eta)+\Bigl(q^{2}-\frac{a''}{a} \Bigr) y_{\lambda}(\bm{q},\eta)=0, \label{eq:eomofmode}
\end{equation}
where $q=|\bm{q}|$. To quantize the tensor perturbation, we impose the canonical commutation relations
\begin{align}
&[\hat{y}_{\lambda} (\bm{q},\eta) , \hat{y}_{\lambda'} (\bm{q}',\eta)]=[\hat{\pi}_{\lambda} (\bm{q},\eta) , \hat{\pi}_{\lambda'} (\bm{q}',\eta)]=0, \label{eq:CCR1} \\
&[\hat{y}_{\lambda} (\bm{q},\eta) , \hat{\pi}_{\lambda'} (\bm{q}',\eta)]=i\delta_{\lambda \lambda'}\delta^{3}(\bm{q}+\bm{q}'). \label{eq:CCR2}
\end{align}
We denote the solution of the equation of motion \eqref{eq:eomofmode} as $f_{q}$ and define the function $g_{q}=i(f'_{q}-a' f_q/a)$. We fix the normalization of the mode function as
\begin{equation}
f_{q}(\eta)g^{*}_{q}(\eta)+f^{*}_{q}(\eta) g_{q}(\eta)=1,  \label{eq:norm}
\end{equation}
and  expand the canonical variables $\hat{y}_\lambda$ and
$\hat{\pi}_\lambda$ as follows:
\begin{align}
\hat{y}_\lambda (\bm{q},\eta)&= f_{q}(\eta) \hat{a}_{\lambda}(\bm{q})+f^{*}_{q}(\eta)\hat{a}^{\dagger}_\lambda (-\bm{q}), \label{eq:expofy} \\
\hat{\pi}_\lambda (\bm{q},\eta)&=(-i) \left( g_{q}(\eta) \hat{a}_{\lambda}(\bm{q})-g^{*}_{q}(\eta) \hat{a}^{\dagger}_\lambda (-\bm{q})\right), \label{eq:expofpi} 
\end{align}
where  $\hat{a}_\lambda$ is the  annihilation operator
satisfying
\begin{align}
&[\hat{a}_{\lambda} (\bm{q}) , \hat{a}_{\lambda'} (\bm{q}')]=0, \label{eq:alg1} \\
&[\hat{a}_{\lambda} (\bm{q}) , \hat{a}^{\dagger}_{\lambda'} (\bm{q}')]=\delta_{\lambda \lambda'}\delta^{3}(\bm{q}-\bm{q}'). \label{eq:alg2}
\end{align}
The equation of the mode function is solved for each epoch, and
junction conditions at $\eta=\eta_\text{r}$ and $\eta=\eta_\text{m}$
yield the full solution of the tensor perturbation in the FLRW
universe. We adopt the following mode function for the
inflation era
\begin{equation}
u^\text{inf}_{q}(\eta)=\frac{1}{\sqrt{2q}} \left( 1- \frac{i}{q(\eta-2\eta_\text{r})} \right)e^{-iq(\eta-2\eta_\text{r})}, \label{eq:BDmode}
\end{equation}
and assume that the initial  quantum
state of PGWs is the Bunch-Davies vacuum $\ket{0^\text{BD}}$ defined
by
\begin{equation}
\hat{a}_\lambda (\bm{q}) \ket{0^\text{BD}}=0. \label{eq:BDvac}
\end{equation}

With the junction conditions, we find the full solution of the mode function as
\begin{align}
  f_{q}(\eta) = 
\begin{cases}
    u^\text{inf}_{q}(\eta) & (-\infty< \eta \leq \eta_\text{r}) \\
    \alpha_{q} u^\text{rad}_{q}(\eta)+\beta_{q} u^{\text{rad}*}_{q}(\eta)  &(\eta_\text{r} < \eta \leq \eta_\text{m}) \\
   \gamma_{q} u^\text{mat}_{q}(\eta)+\delta_{q} u^{\text{mat}*}_{q}(\eta)  &(\eta_\text{m} < \eta )
  \end{cases}, \label{eq:solfq}
\end{align}
where $u^\text{rad}_{q}$ and $u^\text{mat}_{q}$ are the positive
frequency mode solutions in the radiation- and matter-dominated
era and the coefficients $\alpha_q,\beta_q,\gamma_q$ and $\delta_q$
are fixed by the junction conditions. In
particular, the mode function $u^\text{rad}_q$ is given as
\begin{equation}
u^\text{rad}_{q}(\eta)=\frac{1}{\sqrt{2q}}\,e^{-iq\eta}. \label{eq:uR}
\end{equation}
From the solution $f_{q}$, the function $g_{q}$ is obtained as 
\begin{align}
  g_{q}(\eta) = 
\begin{cases}
    v^\text{inf}_{q}(\eta) & (-\infty< \eta \leq \eta_\text{r}) \\
    \alpha_{q} v^\text{rad}_{q}(\eta)-\beta_{q} v^{\text{rad}*}_{q}(\eta)  &(\eta_\text{r} < \eta \leq \eta_\text{m}) \\
   \gamma_{q} v^\text{mat}_{q}(\eta)-\delta_{q} v^{\text{mat}*}_{q}(\eta)  &(\eta_\text{m} < \eta )
  \end{cases}, \label{eq:solgq}
\end{align}
where the functions $v^\text{inf}_q, v^\text{rad}_q$ and $v^\text{mat}_q$
are given by the definition of the function $g_q (\eta)$. The explicit 
formulas of $v^\text{inf}_q$ and $v^\text{rad}_q$ are
\begin{align}
v^\text{inf}_{q}(\eta)&=\sqrt{\frac{q}{2}}\, e^{-iq \eta}, \label{eq:vI} \\
v^\text{rad}_{q}(\eta)&=\sqrt{\frac{q}{2}}\, \Bigl(1-\frac{i}{q\eta} \Bigr)e^{-iq \eta}. \label{eq:vR} 
\end{align}
The normalizations of $u^\text{inf}_{q}, v^\text{inf}_q, u^\text{rad}_q$,
and $v^\text{rad}_q$ are chosen so that Eq.~\eqref{eq:norm} 
is satisfied for each pair
$(u^\text{inf}_q, v^\text{inf}_q)$ and $(u^\text{rad}_q,v^\text{rad}_q)$. The
Bogolyubov coefficients $\alpha_q, \beta_q, \gamma_q$ and $\delta_q$
 satisfy the normalization conditions
\begin{equation}
  |\alpha_{q}|^{2}-|\beta_{q}|^{2}=1,\quad |\gamma_q|^2-|\delta_q|^2=1. \label{eq:normforbogolyubov} 
\end{equation}
The coefficients $\alpha_q$ and $\beta_q$ are determined
by the junction conditions at $\eta=\eta_\text{r}$:
\begin{equation}
\alpha_{q}= \Bigl(1+\frac{i}{q\eta_\text{r}}-\frac{1}{2q^{2} \eta_\text{r}^{2}} \Bigr)e^{2iq\eta_\text{r}}, \quad \beta_{q}= \frac{1}{2q^{2}\eta_\text{r}^{2}}. \label{eq:alphabeta}
\end{equation}
The explicit formulas of the functions $u^\text{mat}_q, v^\text{mat}_q$ and
the coefficients $\gamma_q , \delta_q$ are not needed in the following
analysis. This is because we are interested in the super-horizon mode
at the end of inflation and the sub-horizon mode at the
radiation-matter equality time, that is,
\begin{equation}
q\eta_\text{r} \ll 1, \quad q\eta_\text{m} \gg 1. \label{eq:qobs}
\end{equation}
The sub-horizon condition $q \eta_\text{m} \gg 1$ implies that the
solution $f_{q}$ in the matter era can be approximated
by that for the radiation era.

\begin{figure}[H]
   \centering
   \includegraphics[width=1.00\linewidth]{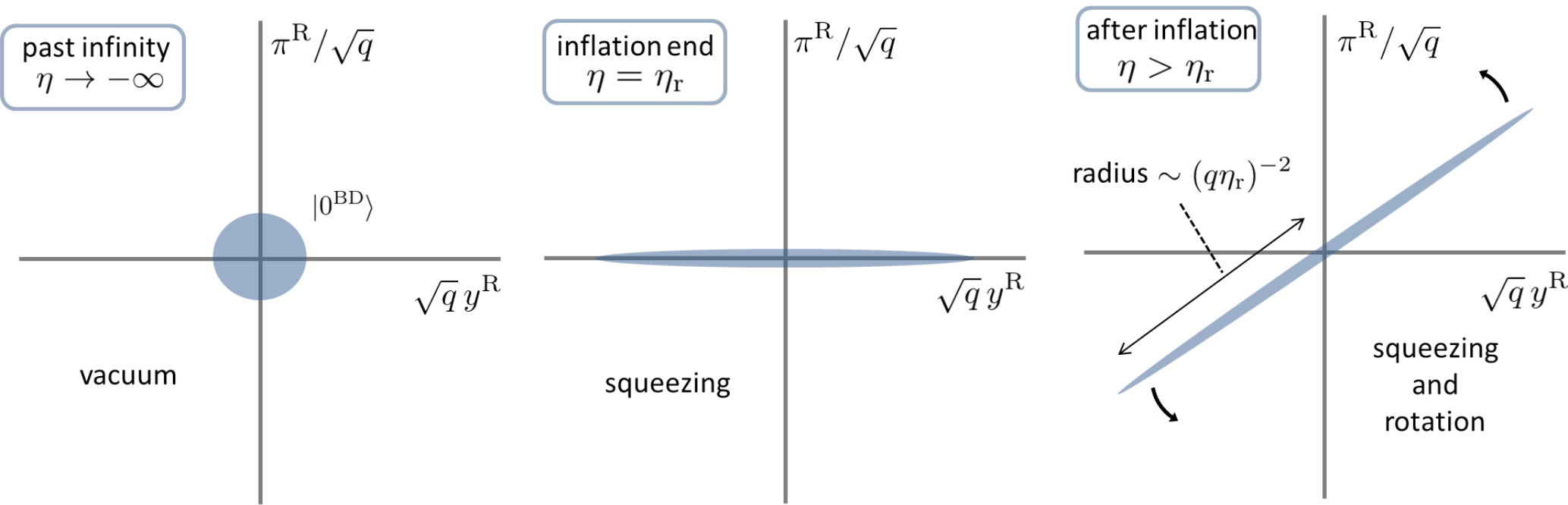}
   \caption{The behavior of the Wigner ellipse of the Bunch-Davis
     vacuum in the phase space
     $(\sqrt{q}\,y^\text{R}, \pi^\text{R}/\sqrt{q})$. Left panel: the
     initial vacuum at the past infinity. Middle panel: the squeezed
     Wigner ellipse at the end of inflation. Right panel: the
     squeezed and rotated ellipse after the inflation.}
    \label{fig:BDbehavior}
\end{figure}

Let us demonstrate the oscillatory behavior of the correlation function of PGWs. 
In order to make a clear connection
between the oscillatory behavior and quantum correlations, we
introduce
\begin{equation}
\hat{A}_\lambda (\bm{q}, \eta) =\sqrt{\frac{q}{2}}\, \hat{y}_{\lambda}(\bm{q},\eta)+\frac{i}{\sqrt{2q}} \hat{\pi}_\lambda (\bm{q},\eta).  \label{eq:A}
\end{equation}
The operator $\hat{A}_\lambda$ for a sub-horizon mode is equivalent to the annihilation operator defined by the positive
frequency mode in each era. In fact, in the radiation or the matter
era $\eta_\text{r} <\eta $, the operator $\hat{A}_\lambda $ for 
the sub-horizon mode $q\eta \gg 1$ is approximated as
\begin{equation}
\hat{A}_\lambda (\bm{q},\eta) \sim \hat{b}_\lambda (\bm{q}) e^{-iq \eta }, \label{eq:approxA}
\end{equation}
where  $\hat{b}_\lambda$ is given by
\begin{equation}
\hat{b}_{\lambda}(\bm{q})=\alpha_{q} \hat{a}_{\lambda}(\bm{q})+\beta^{*}_{q} \hat{a}^{\dagger}_{\lambda}(-\bm{q}). \label{eq:b}
\end{equation}
The operator $\hat{b}_\lambda $ are the annihilation operator defined
by the positive frequency mode $u^\text{rad}_q$ after inflation
($u^\text{rad}_q$ is also the positive frequency mode in the matter era
for $q \eta_\text{m} \gg 1$). Hence the operator $\hat{A}_\lambda$
for the sub-horizon mode has the same role as $\hat{b}_\lambda$.
The correlation function for the field amplitude $\hat{y}_\lambda $ is
\begin{equation}
\bra{0^\text{BD}} \hat{y}_\lambda (\bm{q},\eta) \hat{y}_{\lambda'} (\bm{q}',\eta) \ket{0^\text{BD}}
=\frac{1}{2q} \left(2n_{q}(\eta)+1+c_{q}(\eta)+c^{*}_{q}(\eta)\right) \delta_{\lambda \lambda'} \delta^{3}(\bm{q}+\bm{q}'), \label{eq:yy} 
\end{equation}
where we used
$\hat{y}_{\lambda}(\bm{q},\eta)=(\hat{A}_\lambda(\bm{q},\eta)+\hat{A}^{\dagger}_\lambda
(-\bm{q},\eta))/\sqrt{2q}$
and introduced $n_q$ and $c_q$ by
\begin{align}
  \bra{0^\text{BD}} \hat{A}^{\dagger}_\lambda (\bm{q},\eta) \hat{A}_{\lambda'} (\bm{q}',\eta) \ket{0^\text{BD}}
  &=n_{q}(\eta)  \delta_{\lambda \lambda'}
    \delta^{3}(\bm{q}-\bm{q}'), \label{eq:AdaggerA} \\
\bra{0^\text{BD}} \hat{A}_\lambda (\bm{q},\eta) \hat{A}_{\lambda'} (\bm{q}',\eta) \ket{0^\text{BD}}
&=c_{q}(\eta)  \delta_{\lambda \lambda'} \delta^{3}(\bm{q}+\bm{q}'). \label{eq:AA}
\end{align}
The function $n_q$ represents the mean particle number 
and $c_q$ characterizes the quantum coherence of the Bunch-Davies
vacuum. The functions $n_q$ and $c_q$ completely determine the
quantum property of the Bunch-Davies vacuum.  We evaluate the correlation 
function in the matter era. For the target range of the wave number
$1/\eta_\text{m} \ll q \ll 1/\eta_\text{r}$ \eqref{eq:qobs}, the
functions $n_q$ and $c_q$ for the sub-horizon mode $q\eta \gg 1$ are computed as
\begin{align}
  n_{q}(\eta) & \sim |\beta_q|^{2}, \label{eq:approxnq} \\
c_q(\eta) & \sim \alpha_q \beta^{*}_q e^{-2iq\eta} \sim -|\beta_q |^{2} 
                 e^{-2iq\eta}, \label{eq:approxcq}
\end{align}
where the second approximation in Eq.~\eqref{eq:approxcq}
follows from $q\eta_\text{r} \ll 1$.  The behavior of the correlation function of $\hat{y}_\lambda$ in the matter-dominated era is
  obtained as
\begin{equation}
\bra{0^\text{BD}}\hat{y}_\lambda (\bm{q},\eta) \hat{y}_{\lambda'} (\bm{q}',\eta) \ket{0^\text{BD}}
\sim \frac{|\beta_q|^{2}}{q} (1-\cos(2q\eta) ) \delta_{\lambda \lambda'} \delta^{3}(\bm{q}+\bm{q}'), \label{eq:approxyy}
\end{equation}
where the cosine term comes from $c_q(\eta)$, and the correlation function oscillates in time. In terms of the Fock space defined by
$\hat{A}_\lambda$, the Bunch-Davies vacuum can be expressed as
\begin{align}
\ket{0^\text{BD}}
&\propto \bigotimes_{\bm{q} \in \mathbb{R}^{3+}}
  \bigotimes_{\lambda} \exp \left[\frac{c_q}{n_q +1}
  \hat{A}^{\dagger}_\lambda(\bm{q},\eta)\hat{A}^{\dagger}_\lambda(-\bm{q},\eta)
  \right] \ket{0;\eta} \nonumber \\
&= \bigotimes_{\bm{q} \in \mathbb{R}^{3+}}  \bigotimes_{\lambda}  \sum^{\infty}_{n=0} \left(\frac{c_q}{n_q +1} \right)^{n} \ket{n_{\bm{q},\lambda} \, n_{-\bm{q},\lambda}; \eta} , \label{eq:BDinAFock}
\end{align}
where the state $\ket{0;\eta}$ is defined by
$\hat{A}_{\lambda}(\bm{q},\eta)\ket{0;\eta}=0$ and
$\mathbb{R}^{3+}:=\{(x,y,z)| (x,y,z) \in \mathbb{R}^{3}, z \geq 0
\}$.
The function $c_q$, which characterizes the coherence between the
modes $\bm{q}$ and $-\bm{q}$, leads to the squeezing and rotation of
the Wigner function in the phase space. From Eq.~\eqref{eq:BDvac}, 
the wave function of the Bunch-Davies vacuum for a
single mode $\bm{q}$ and a polarization $\lambda$ is 
\begin{equation}
\psi^\text{BD}(y, \eta)=\sqrt{\frac{2\Omega_q^\text{R}}{\pi}} \exp(-\Omega_q (\eta)|y|^{2}), \quad \Omega_q (\eta)= \frac{g^{*}_q(\eta)}{f^{*}_q (\eta)}, \label{eq:psiBD} 
\end{equation}
where we omitted the labels $\bm{q}$ and $\lambda$, and the superscript R denotes the real part. The Wigner function
$W^\text{BD}(y, \pi_{y} ,\eta)$ of the density matrix 
$\rho^\text{BD}(y,y',\eta)=\psi^\text{BD}(y, \eta)
[\psi^\text{BD}(y',\eta)]^{*}$ is given by
\begin{align}
W^\text{BD}(y,\pi_{y},\eta)
&=\frac{1}{(2\pi)^{2}} \int dx^\text{R} dx^\text{I} \,  e^{i( \pi^\text{R}_{y} x^\text{R}+\pi^\text{I}_{y} x^\text{I})} \rho^\text{BD} (y-x/2, y+x/2,\eta) \nonumber \\
&=w^\text{BD} (y^\text{R}, \pi^\text{R}_{y},\eta) w^\text{BD} (y^\text{I}, \pi^\text{I}_{y},\eta), \label{eq:Wigner} \\
w^\text{BD}(x, p,\eta)&= \frac{1}{\pi } \exp \left[ -2\Omega_q^\text{R} \, x^{2} -\frac{2}{\Omega^\text{R}_q} \left(p +\Omega_{q}^\text{I} \, x\right)^{2} \right], \label{eq:wigner}
\end{align}
where the superscript I denotes the imaginary part. 
Fig.~\ref{fig:BDbehavior} schematically
represents the behavior of the Wigner function
$w^\text{BD}(y^\text{R}, \pi^\text{R}_y, \eta)$.
\noindent
In Fig.~\ref{fig:BDbehavior}, the left panel represents the initial
vacuum state at the past infinity $\eta \rightarrow -\infty$ and the
middle panel represents the squeezed vacuum by the inflationary
expansion. The right panel shows the Wigner ellipse after the end of 
inflation for a sub-horizon mode. The Wigner function is further squeezed until
the horizon re-entry. After that, the Wigner ellipse rotates during
the radiation and matter era. (Its thickness is around $\hbar=1$ in
the right panel of Fig.~\ref{fig:BDbehavior}, however it can be
ignored in \eqref{eq:approxyy}.) The oscillation of the correlation function
corresponds to the rotation of the Wigner ellipse in the phase space.

In order to understand the oscillatory feature from the viewpoint of
quantum superpositions, we have introduced the two modes $\bm{q}$
and $-\bm{q}$ by defining the annihilation operator
\eqref{eq:A}.  On the other hand, we have used the Wigner function of
the single mode $\bm{q}$ for the real (or imaginary) part of
the field $\hat{y}_\lambda$ to explain the squeezing feature of the
state. These two treatments are connected by the following relation
\begin{equation}
\hat{y}^\text{R}_{\lambda}(\bm{q},\eta)
=\frac{1}{2\sqrt{2q}}\left(\hat{A}_{\lambda} (\bm{q},\eta)+\hat{A}^\dagger_{\lambda} (-\bm{q},\eta)+\hat{A}^\dagger_{\lambda} (\bm{q},\eta)+\hat{A}_{\lambda} (-\bm{q},\eta)\right), \label{eq:yR}
\end{equation}
where
$\bm{q} \in \mathbb{R}^{3+}$
because of the relation
$\hat{y}^\text{R}_\lambda(-\bm{q},\eta)=\hat{y}^\text{R}_\lambda
(\bm{q},\eta)$.
The correlation function of  $y^\text{R}_\lambda$ is 
\begin{equation}
\bra{0^\text{BD}} \hat{y}^\text{R}_\lambda
(\bm{q},\eta)\hat{y}^\text{R}_{\lambda'} (\bm{q}',\eta)
\ket{0^\text{BD}}=
\frac{1}{4q} \left(2n_q (\eta) +1 +c_q (\eta) +c^{*}_q (\eta)\right) \delta_{\lambda \lambda'}\delta^{3}(\bm{q}-\bm{q}'), \label{eq:yRyR}
\end{equation}
and contains the function $c_q(\eta)$ characterizing the
quantum coherence for the modes $\bm{q}$ and $-\bm{q}$.

%
\section{Relation between the oscillatory behavior and quantum discord}
\label{sec:discord}

In this section we clarify
the relation between the oscillatory behavior of the CMB angular-power
  spectrum caused by PGWs and quantum discord. For this purpose, we
introduce the notion of the \textit{classically correlated state}. A
given bipartite state $\rho_\text{AB}$ is called classically
correlated \cite{Oliver2001, Oppenheim2002} if the state has the
following form
\begin{equation}
\rho_\text{AB}= \sum_{i,j} p_{ij} \ket{\psi^{i}_\text{A}} \bra{\psi^{i}_\text{A}} \otimes \ket{\phi^{j}_\text{B}} \bra{\phi^{j}_\text{B}}, \label{eq:NDstate}
\end{equation}
where $p_{ij}$ is a joint probability 
  ($p_{ij} \geq 0, ~\sum_{i,j}p_{ij}=1$) and characterizes the classical
  correlation between A and B. The vectors $\ket{\psi^{i}_\text{A}} $
and $ \ket{\phi^{k}_\text{B}}$ of each system A and B satisfy the
orthonormal conditions
\begin{equation}
\langle \psi^{i}_\text{A}| \psi^{j}_\text{A} \rangle=\delta^{ij}, \quad \langle \phi^{k}_\text{B}| \phi^{l}_\text{B} \rangle=\delta^{kl}. \label{eq:orthonormalpsiphi}
\end{equation}
The particular feature of classically correlated states is that there
is a rank-1 projective measurement for the subsystem A or B such that
the states are not disturbed \cite{Oliver2001} in the following sense:
\begin{equation}
\sum_{i} \hat{P}^i_\text{A} \rho_\text{AB} \hat{P}^{i}_\text{A} =\sum_{j} \hat{P}^j_\text{B} \rho_\text{AB} \hat{P}^{j}_\text{B}=\rho_\text{AB}, \label{eq:nondisturb}
\end{equation}
where $\hat{P}^{i}_\text{A}$ and $\hat{P}^{j}_\text{B}$ are rank-1
projective operators satisfying
$\sum_{i}\hat{P}^{i}_\text{A}=\hat{\mathbb{I}}_\text{A}$ and
$\sum_{i}\hat{P}^{i}_\text{B}=\hat{\mathbb{I}}_\text{B}$. This
property is not required for separable states (non-entangled states)
\cite{Werner1989} defined by
\begin{equation}
\rho_\text{AB}= \sum_{i} \lambda_{i} \rho^{i}_\text{A}  \otimes \sigma^{i}_\text{B}, \label{eq:sepstate}
\end{equation}
where $\lambda_i$ is a probability, and $\rho^{i}_\text{A}$ and
$\sigma^{i}_\text{B}$ are density operators. This is because
$\rho^{i}_\text{A}$ and $\rho^{j}_\text{A}$ ($i\neq j$) do not
have to commute each other generally, and hence separable
states can be disturbed by a projective measurement for the subsystem
A. It is obvious that the classically correlated states are
included in the separable states by the definitions of each state.

Next we introduce quantum discord \cite{Oliver2001} as a measure of
 quantum correlations. Quantum discord is the difference between the mutual
information of a given bipartite state $\rho_\text{AB}$ and its
generalization with a projective measurement. The mutual
information $I_\text{AB}$ is
\begin{equation}
I_\text{AB}=S_\text{A}+S_\text{B}-S_\text{AB}, \label{eq:I}
\end{equation}
where $S_\text{A}, S_\text{B}$ and $S_\text{AB}$ are the von Neumann
entropy of the density operators
$\rho_\text{A}=\text{Tr}_\text{B} [\rho_\text{AB}]$,
$\rho_\text{B}=\text{Tr}_\text{A} [\rho_\text{AB}]$ and
$\rho_\text{AB}$, respectively. For example,
$S_\text{A}=S(\rho_\text{A})=-\text{Tr}_\text{A}[\rho_\text{A} \log
\rho_\text{A}]$.
The mutual information characterizes the total correlation of the
bipartite state $\rho_\text{AB}$. Using the conditional entropy
$S_{\text{B}|\text{A}}=S_\text{AB}-S_\text{A}$, the mutual information
is rewritten as
\begin{equation}
I_\text{AB}=S_\text{B}-S_{\text{B}|\text{A}}. \label{eq:I1}
\end{equation}
This second expression leads to the notion
of quantum discord. As a generalization of the conditional
entropy with a projective measurement, we can consider 
\begin{equation}
J_{\text{B}|\{ \hat{P}^{j}_\text{A}\}}=S_\text{B}-\sum_{i} p_{i} S_{\text{B}|\hat{P}^{i}_\text{A}}, \label{eq:J}
\end{equation}
where
$p_i =\text{Tr}_\text{AB}[\hat{P}^{i}_\text{A} \rho_\text{AB}]$
and $S_{\text{B}|\hat{P}^{i}_\text{A}}$ is the von Neumann entropy of
the density operator given by
\begin{equation}
\rho^{i}_\text{B}=\frac{\text{Tr}_\text{A}[\hat{P}^{i}_\text{A} \rho_\text{AB} \hat{P}^{i}_\text{A}]}{p_i}. \label{eq:rhoiB}
\end{equation}
The von Neumann entropy
$\sum_{i} p_{i} S_{\text{B}|\hat{P}^{i}_\text{A}}$ is equivalent to
the conditional entropy after the projective measurement
$\hat{P}^{i}_\text{A}$ on the system A. Quantum discord of a bipartite
state $\rho_\text{AB}$ is the minimum of difference between the two
mutual informations:
\begin{equation}
\delta_{\text{B}|\text{A}}:=I_\text{AB}-\max_{\hat{P}^{j}_\text{A}} J_{\text{B}|\{ \hat{P}^{j}_\text{A} \}}, \label{eq:discord} 
\end{equation}
where we maximize over all possible projective measurements on the
system A. In general, $\delta_{\text{B}|\text{A}}$ is not the
  same as $\delta_{\text{A}|\text{B}}$. In Ref.~\cite{Oliver2001}, it was
shown that $\delta_{\text{B}|\text{A}}=0=\delta_{\text{A}|\text{B}}$
for a given bipartite state if and only if the state is classically
correlated. The quantities $\delta_{\text{B}|\text{A}}$ and
$\delta_{\text{A}|\text{B}}$ are good indicators of the quantumness of
the correlation associated with a given state.

Now, we construct a classical model (zero quantum discord
  state) of PGWs. Firstly, we impose the following three assumptions on the
classical model:
\begin{description}
  \item[Assumption 1.] The mode obeys the linearized Einstein equation.  
  \item[Assumption 2.] The initial state is a Gaussian state.
  \item[Assumption 3.] The initial state is invariant under spatial translations and rotations.    
\end{description}
These assumptions are accepted in the standard treatment of the linear
quantum fluctuations in the FLRW universe. We denote the
classical model (state) of PGWs as $\rho^\text{cl}$. By the
assumption 1, the evolution of the Heisenberg operators is determined
and hence we only have to fix the initial condition of the
state $\rho^\text{cl}$ to identify the classical model. From the
assumptions 2 and 3, the state $\rho^\text{cl}$ has the following
expectation values for $\hat{b}_\lambda$ and
$\hat{b}^\dagger_\lambda$ defined by \eqref{eq:b} :
\begin{align}
&\text{Tr}[\hat{b}_\lambda (\bm{q}) \rho^\text{cl}]=0, \label{eq:expb} \\
&\text{Tr} [ \hat{b}^{\dagger}_\lambda (\bm{q}) \hat{b}_{\lambda'} (\bm{q}') \rho^\text{cl}]=m_{q}\,  \delta_{\lambda \lambda'} \delta^{3}(\bm{q}-\bm{q}'), \label{eq:mq} \\
&\text{Tr}[ \hat{b}_\lambda (\bm{q}) \hat{b}_{\lambda'} (\bm{q}') \rho^\text{cl}]=d_q \,\delta_{\lambda \lambda'} \delta^{3}(\bm{q}+\bm{q}') , \label{eq:dq}
\end{align}
where $m_{q}$ and
$d_{q}$ are free functions characterizing the initial state. Because of the
translational invariance, the expectation value of the annihilation
operator $\hat{b}_\lambda$ with nonzero modes vanishes.  From the
assumption of being Gaussian state, the functions $m_q$ and $d_q$
completely determine the form of the state $\rho^\text{cl}$.

In order to fix the two functions $m_q$ and $d_q$, we further impose the following two assumptions:
\begin{description}
  \item[Assumption 4.]  The bipartite state with modes $\bm{q}$
  and $-\bm{q}$ defined by the annihilation and creation operators
  $\hat{b}_\lambda (\bm{q})$ and $\hat{b}^\dagger_\lambda(\bm{q})$ is
  a classically correlated state (zero discord state).
  \item[Assumption 5.] At the present time, the classical model provides the
  same correlation function of PGWs as the initial Bunch-Davies vacuum.
\end{description}
From the assumption 2,3 and 4, we can find that the state $\rho^\text{cl}$
is classically correlated if and only if the function
$d_q$ vanishes. Let us show this statement. For simplicity, we omit the
index of the polarization $\lambda$ and denote the state $\rho^\text{cl}$
with the mode $\bm{q}$ and $-\bm{q}$ as
$\rho^\text{cl}_{\bm{q},-\bm{q}}$. When the function $d_q$ vanishes,
the Gaussian state $\rho_{\bm{q},-\bm{q}}$ is a product state, which corresponds to a classically correlated state with the weight 
$p_{ij}=p_i^{\text{A}}p_j^{\text{B}}$ in Eq.~\eqref{eq:NDstate}.
Conversely, if the state
$\rho^\text{cl}_{\bm{q},-\bm{q}}$ is classically correlated, then the
state $\rho^\text{cl}_{\bm{q},-\bm{q}}$ is represented by a product
state
\begin{equation}
\rho^\text{cl}_{\bm{q},-\bm{q}}=  \rho_{\bm{q}}  \otimes \sigma_{-\bm{q}}, \label{eq:prodq-q}
\end{equation}
where $\rho_{\bm{q}}$ and $\sigma_{-\bm{q}}$ are density operators for
each mode. In general, a given classically correlated state can have
correlation, but classically correlated Gaussian states are
product states \cite{Giorda2010, Adesso2010}. The Appendix A is
devoted to a simple proof of this property. Then the expectation value
of $\hat{b} (\bm{q}) \hat{b} (-\bm{q})$ is given by
\begin{equation}
\text{Tr}[\hat{b} (\bm{q}) \hat{b} (-\bm{q}) \, \rho^\text{cl}_{\bm{q},-\bm{q}} ]= \text{Tr}[\hat{b} (\bm{q}) \, \rho_{\bm{q}} ]  \times \text{Tr}[\hat{b} (-\bm{q}) \, \sigma_{-\bm{q}}]=0, 
\end{equation}
because the one-point function of the annihilation operator
$\hat{b}(\bm{q})$ vanishes by the translation invariance
\eqref{eq:expb}. Hence the function $d_q$ must vanish. As $d_q$
characterizes the coherence of $\rho^\text{cl}$ (see Eq.~\eqref{eq:dq}), 
the following statement holds: the quantum discord
exists if and only if the quantum coherence for the modes $\bm{q}$
and $-\bm{q}$ exists.

We emphasize that the condition $d_q=0$ for the classical state
cannot be derived from the
separability. To judge
whether a given bipartite state $\rho_\text{AB}$ is entangled or not,
the positive partial transposed (PPT) criterion is useful
\cite{Peres1996, Horodecki1996}; if a bipartite state $\rho_\text{AB}$
is separable then the following inequality holds
\begin{equation}
(\rho_\text{AB})^{\text{T}_\text{B}} \geq 0, \label{eq:PPT}
\end{equation}
where $\text{T}_\text{B}$ is the transposition for the
subsystem B and the inequality means that
$(\rho_\text{AB})^\text{T}_\text{B}$ has no negative
eigenvalues. For the Gaussian
bipartite state $\rho^\text{cl}_{\bm{q},-\bm{q}}$ defined by
$\hat{b}(\bm{q})$ and $\hat{b}(-\bm{q})$, it is known that the PPT
criterion is the necessary and sufficient condition for the
separability \cite{Simon2000, Giedke2001, Fujikawa2009}. The inequality
\eqref{eq:PPT} for the state $\rho^\text{cl}_{\bm{q},-\bm{q}}$ is given by
\begin{equation}
m_q \geq |d_q|. \label{eq:PPT2}
\end{equation}
The derivation of the inequality \eqref{eq:PPT2} is shown in the
appendix B. We can admit the non-entangled model of PGWs with nonzero
$d_q$ (non-zero discord). Such a model has the following expectation value for the sub-horizon modes ($q\eta \gg 1$),
\begin{equation}
\text{Tr}[\hat{A}_\lambda (\bm{q}, \eta) \hat{A}_{\lambda'} (\bm{q}', \eta) \rho^\text{cl}] \sim d_q e^{-2iq\eta} \delta_{\lambda \lambda'} \delta^{3}(q+q'), \label{eq:AAcl}
\end{equation}
and shows the oscillatory behavior of the correlation function. Hence we cannot
distinguish whether the model has quantum entanglement (between
$\bm{q}$ and $-\bm{q}$ modes) by just observing the oscillatory
behavior.

The function $m_q$ is determined by the assumption 5. Using the
approximated form of the annihilation operator $\hat{A}_\lambda$ for
the sub-horizon scale \eqref{eq:approxA}, we obtain the
correlation function of the state $\rho^\text{cl}$ for $q\eta_0 \gg 1$ as
\begin{equation}
\text{Tr}[\hat{y}_\lambda(\bm{q},\eta_0) \hat{y}_{\lambda'}
(\bm{q}',\eta_0) \rho^\text{cl}] \sim \frac{1}{2q}(2m_q+1)
\delta_{\lambda \lambda'} \delta^{3}(\bm{q}+\bm{q}'), \label{eq:poweryyforND} 
\end{equation}
where $\eta_0$ is the conformal time of the present day. 
The assumption 5 requires that the correlation function of the variables
$\hat{y}_\lambda$ should be equal to that given by the Bunch-Davies
vacuum \eqref{eq:approxyy}. For $q \eta_0 \gg 1$ and $q\eta_\text{r} \ll 1$ the function $m_q $ can be fixed as
\begin{equation}
m_q= n_q (\eta_0) +\frac{1}{2}\Bigl(c_q (\eta_0)+c^{*}_q(\eta_0) \Bigr) \sim  |\beta_q|^{2} (1-\cos(2q\eta_0)), \label{eq:explicitmq}
\end{equation}
where we used Eq.~\eqref{eq:approxyy} at the present time $\eta_0$.

Here we compare our analysis with the previous work
\cite{Bose2002}. They considered squeezed and non-squeezed models of
PGWs. Both of these models assume the Bunch-Davies vacuum as
the initial state of PGWs. The squeezed model corresponds to PGWs
treated in the previous section. The non-squeezed one is constructed
by assuming the following form of the mode function in the
matter-dominated era
\begin{equation}
f_{q}(\eta) \propto\frac{e^{-iq\eta}}{\sqrt{2q}}, \label{eq:travelingwave}
\end{equation}
which has only the positive frequency mode. This means that there is
no particle production and any squeezing effects. In \cite{Bose2002},
the specification \eqref{eq:travelingwave} of the mode function was
called the traveling wave condition, which corresponds to the
classically correlated assumption in our analysis. The amplitude
of the mode \eqref{eq:travelingwave} is determined by the same
procedure as our assumption 5, which was called the fair comparison in
\cite{Bose2002}. For the sub-horizon mode at the present time
$q\eta_0 \gg 1$, the amplitude was given by $m_q$ without the
cosine term in \cite{Bose2002}. The disregard of the cosine term is valid in the
calculation of the angular-power spectrum. We will
explain the detail of this statement later (after
Eq.~\eqref{eq:Clcl2}).

Let us compare the two models of PGWs by the angular-power spectrum of
CMB temperature fluctuations. The temperature fluctuations caused by
the tensor perturbation is
\begin{equation}
\frac{\delta \hat{T} (\hat{n})}{T_0}=-\frac{1}{2} \hat{n}_i \hat{n}_j \int^{\eta_0 }_{\eta_\text{L}} d\eta \left[\frac{\partial}{\partial \eta} \hat{h}_{ij} \right]_{r=\eta_0 -\eta}=-\frac{1}{2} \hat{n}_i \hat{n}_j \int^{\eta_0 }_{\eta_\text{L}} \frac{d\eta}{a(\eta)}  \hat{\pi}_{ij} \Bigr|_{r=\eta_0 -\eta}, \label{eq:deltaT}
\end{equation}
where $\hat{n}_i$ is the unit vector describing the direction of CMB
photons' propagation and the CMB photons are emitted at the conformal
time $\eta_\text{L}$. The angular-power spectrum $C_\ell$ is defined by
\begin{equation}
C_\ell=\frac{1}{4\pi} \int d^{2}\hat{n}\,d^{2}\hat{n}' \, P_{\ell}(\hat{n} \cdot \hat{n}') \left\langle{ \frac{\delta \hat{T} (\hat{n})}{T_0} \frac{\delta \hat{T} (\hat{n}')}{T_0}} \right\rangle, \label{eq:Cl}
\end{equation}
where $P_\ell (\hat{n} \cdot \hat{n}')$ is the Legendre polynomial of
degree $\ell$ and the bracket means the expectation value for a state. The
angular-power spectrum for each multipole $\ell$ is characterized by the
redshift factor of the end of inflation $z_\text{end}$, matter-radiation
equality $z_\text{eq}$, the last scattering surface $z_\text{L}$ and
the amplitude of PGWs given by $H_\text{inf}/M_\text{pl}$. We suppose
that the redshift factors are
\begin{align}
1+z_\text{end}&=\frac{a(\eta_0)}{a(\eta_\text{r})}=\frac{(\eta_0+\eta_\text{m})^{2}}{4\eta_\text{m} \eta_\text{r}} \gtrsim 10^{27}, \label{eq:zend} \\
1+z_\text{eq}&=\frac{a(\eta_0)}{a(\eta_\text{m})}=\frac{1}{4}\Bigl(\frac{\eta_0}{\eta_\text{m}}+1\Bigr)^{2} \sim 10^{4}, \label{eq:zeq} \\
1+z_\text{L}&=\frac{a(\eta_0)}{a(\eta_\text{L})}=\Bigl(\frac{\eta_0 + \eta_\text{m}}{\eta_\text{L}+ \eta_\text{m}}\Bigr)^{2} \sim 10^{3}, \label{eq:zL}
\end{align}
where $z_\text{end}$ is estimated for the GUT scale
$H_\text{inf} \sim 10^{15}$ GeV, the present Hubble
$H_0 \sim 10^{-43} $ GeV and the e-folding $N \sim 70$ to solve the
horizon and flatness problem. In the following, we focus on the target
frequency $1/\eta_\text{m} \ll q \ll 1/\eta_\text{r}$. By the
  condition $q\eta_\text{m} \gg 1$, we can use the mode solution in
  the radiation era for the CMB power spectrum. Then we
obtain the following formulas of the angular-power spectrum for 
$\rho^\text{BD}=\ket{0^\text{BD}}\bra{0^\text{BD}}$ and
$\rho^\text{cl}$
\begin{align}
C^\text{BD}_{\ell}
&= \frac{8\pi}{2\ell+1} \int^{\infty}_{0} dq\, q^{2}  \Bigl[(2|\beta_q|^{2}+1) |V_\ell (q)|^{2}-\alpha_q \beta^{*}_q V^{2}_{\ell}(q)- \alpha^{*}_q \beta_q V^{*2}_{\ell}(q) \Bigr], \label{eq:ClBD}  \\
C^\text{cl}_{\ell}
&=\frac{8\pi}{2\ell+1} \int^{\infty}_{0} dq\, q^{2} \, (2m_q +1) |V_\ell (q)|^{2}, \label{eq:Clcl}
\end{align}
where $\alpha_q, \beta_q$ are the Bogolyubov coefficients
\eqref{eq:alphabeta}. The function $V_\ell (q)$ is defined by
\begin{equation}
V_\ell (q)=\frac{\sqrt{2}}{M_\text{pl}} \frac{1}{(2\pi)^{3/2}} \sqrt{\frac{\pi(2\ell+1)(\ell+2)!}{2(\ell-2)!}} \int^{\eta_0}_{\eta_\text{L}} \frac{d\eta}{a(\eta)} \frac{j_{\ell}(q(\eta_0 -\eta))}{q^{2}(\eta_0 - \eta)^{2}} v^\text{rad}_{q} (\eta), \label{eq:Vl} 
\end{equation}
where $j_\ell (z)$ is the spherical Bessel function and $v^\text{rad}_q$
is the positive frequency mode in the radiation era
(Eq.~\eqref{eq:vR}). As the leading order contribution for
  $q\eta_\text{r} \ll 1$, we obtain 
\begin{align}
C^\text{BD}_{\ell}
&\sim \frac{16\pi}{2\ell+1} \int^{\infty}_{0} dq\, q^{2} \, |\beta_q|^2 \Bigl[ |V_\ell (q)|^{2}+ V^{2}_{\ell}(q)/2+ V^{*2}_{\ell}(q)/2 \Bigr], \label{eq:ClBD2}  \\
C^\text{cl}_{\ell}
&\sim \frac{16\pi}{2\ell+1} \int^{\infty}_{0} dq\, q^{2} \, |\beta_q|^{2}(1-\cos[2q\eta_0] ) |V_\ell (q)|^{2}  \nonumber \\
&\sim \frac{16\pi}{2\ell+1} \int^{\infty}_{0} dq\, q^{2} \, |\beta_q|^{2} |V_\ell (q)|^{2}, \label{eq:Clcl2}
\end{align}
where the formula of $m_q$ \eqref{eq:explicitmq} was substituted into \eqref{eq:Clcl} and
the approximations $\alpha_q \sim -\beta_q$ and
$|\beta_q|^{2} +1/2 \sim |\beta_q|^{2}$ were used in the first line of 
\eqref{eq:ClBD2} and \eqref{eq:Clcl2}. In the second
approximation of Eq.~\eqref{eq:Clcl2}, we used the fact that
the cosine term $\cos[2q\eta_0]$ does not contribute to the
$q$-integral  because the present time $\eta_0$ is much larger
than  $\eta_\text{r}, \eta_\text{m}, \eta_\text{L}$ and
the cosine term oscillates rapidly in the integration.

\begin{figure}[H]
   \centering
   \includegraphics[width=0.60\linewidth]{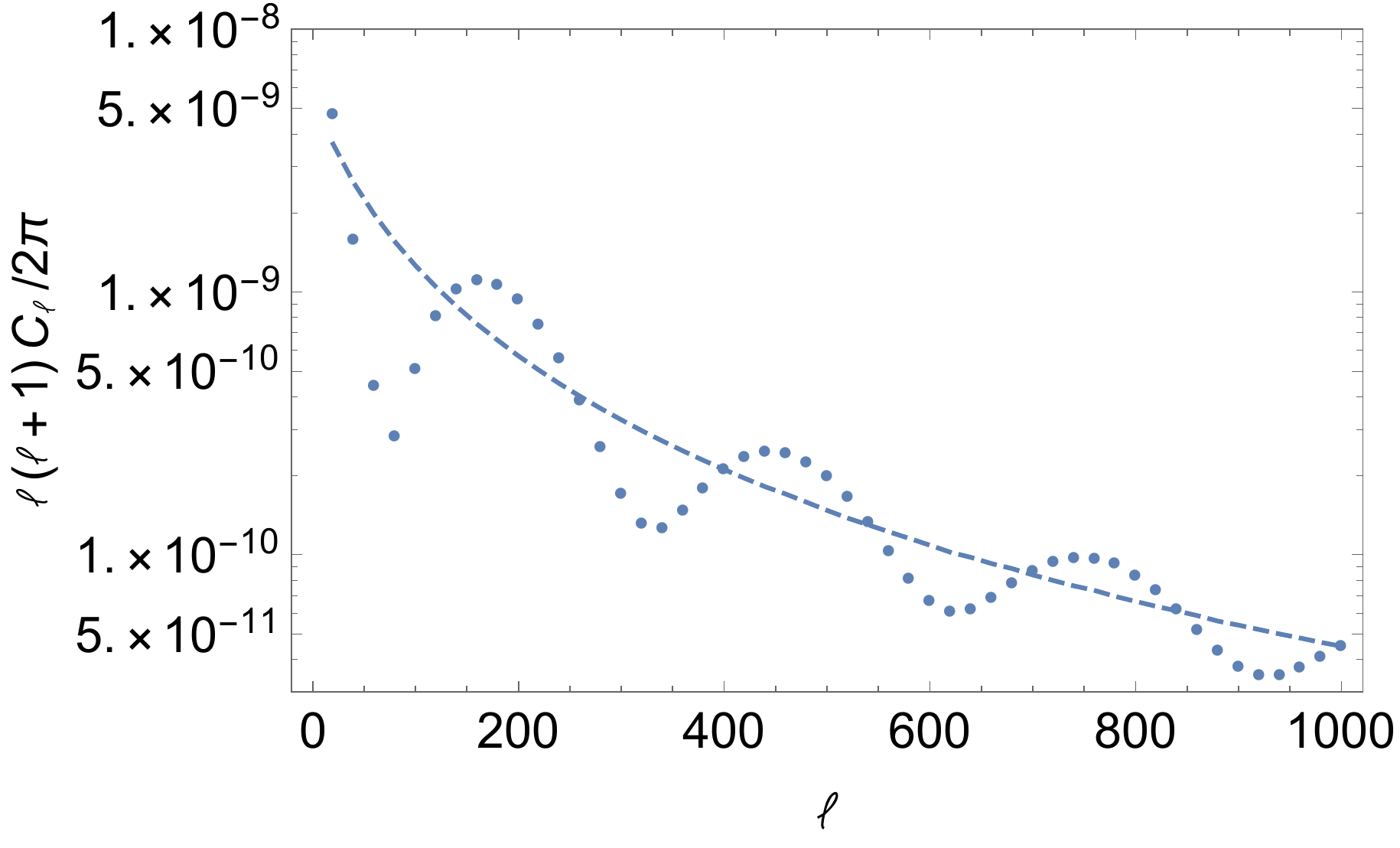}
\caption{The behavior of the angular-power spectrum of CMB temperature
  fluctuations $C^\text{BD}_\ell$ (dotted line) and
    $C^\text{cl}_\ell$ (dashed line). $C_\ell^{\text{BD}}$ shows
    the oscillatory behavior and  $C^\text{cl}_\ell$ does not have such a behavior.}
    \label{fig:BDvscl}
\end{figure}

Fig.~\ref{fig:BDvscl} presents the angular-power spectrum
$C^\text{BD}_\ell$ and $C^\text{cl}_\ell$ given by \eqref{eq:ClBD2}
and \eqref{eq:Clcl2}. $C^\text{BD}_\ell$ shows oscillation, on the
other hand, $C^\text{cl}_\ell$ decreases monotonically as the
multipole $\ell$ increases. The oscillation is attributed to
  the phase factor of $v^\text{rad}_{q}\sim \sqrt{q/2} e^{-iq\eta}$ contained in the last two terms of Eq.~\eqref{eq:ClBD2}. From the redshift factors given
  by \eqref{eq:zend}, $\eqref{eq:zeq}$ and $\eqref{eq:zL}$, the
  typical value of the phase is estimated as follows:
\begin{equation}
 q\eta_\text{L} \sim \frac{ \ell \eta_\text{L}}{\eta_0 -\eta_\text{L}} \sim \frac{\ell}{100}, \label{eq:estphase}
\end{equation}
where we have used $q \sim \ell/(\eta_\text{L}-\eta_0)$. The
oscillation begins from $\ell \sim 100$ (the corresponding phase is $q\,\eta_\text{L} \sim 1$) and the period
of the oscillation is about $100$ up to a numerical factor, which is
observed in Fig.~\ref{fig:BDvscl}.

Let us discuss how a model with free functions $m_q$ and $d_q$ defined in Eqs.~\eqref{eq:mq} and \eqref{eq:dq} shows the oscillatory feature. The formula of the angular-power spectrum for $q\eta_\text{m} \gg 1$ is written by these functions as 
\begin{equation}
C_{\ell}= \frac{8\pi}{2\ell+1} \int^{\infty}_{0} dq\, q^{2} \Bigl[ (2m_q +1) |V_\ell (q)|^{2}-|d_q| \bigl( e^{i\theta_q} V^{2}_{\ell}(q)+e^{-i\theta_q} V^{*2}_{\ell}(q) \bigr) \Bigr],
\label{eq:tildeCl} 
\end{equation}
where $d_q =|d_q|e^{i\theta_q}$ and $V_\ell (q)$ is given by \eqref{eq:Vl}. The second term of the integrand in \eqref{eq:tildeCl} is crucial for the oscillatory feature. If the condition $m_q \gg |d_q| $ holds then the second term is negligible. Choosing $m_q$ as Eq.~\eqref{eq:explicitmq}, we can get the almost same angular-power spectrum as that for the classical state. Also if the phase $\theta_q$ changes rapidly and takes various values in the $q$-integral, then the second term is neglected again by the Riemann-Lebesgue lemma. The PGWs superposed with many phases (the function $d_q$ controls the coherence of PGWs) contribute to the power spectrum, and the oscillation is reduced as a result. For the two situations $m_q \gg |d_q|$ or rapidly changing phase $\theta_q$, the oscillation degrades sufficiently even if the state has nonzero $d_q$, that is, nonzero discord. Therefore we can only conclude that the CMB power spectrum computed from the classical state has no oscillation. The converse statement that the absence of the oscillation means zero quantum discord does not necessarily hold. 

The whole analysis is based on the free theory of the tensor perturbation, 
and the nonlinear interaction with other fields is not included. Since such nonlinear
interactions can induce quantum decoherence generally, there is the
possibility of loss of the quantum feature for PGWs. We discuss the
decoherence effect for the tensor perturbation in the next
section.
%

\section{Decoherence for super-horizon modes and quantum correlations}
\label{sec:decoherence}
Quantum decoherence is the loss of quantum superposition and induced
by the interaction with an environment. In cosmological
  situations, quantum decoherence plays a crucial role to explain
quantum-to-classical transition of primordial fluctuations. In
\cite{Kiefer2007}, the authors discussed the decoherence for primordial
fluctuations with the super-horizon modes and introduced the two
conditions: the decoherence condition and the correlation
condition. In this section, we clarify the meaning of these two
conditions in terms of quantum correlations.

To get a clear intuition of the
decoherence effect, we construct a decohered Gaussian state of
PGWs. We consider the total system with the full Hamiltonian
\begin{equation}
\hat{H}(\eta)=\hat{H}^{y}_{0}(\eta)+\hat{H}^{\varphi}_{0}(\eta)+\hat{V}(\eta), \label{eq:H} 
\end{equation}
where $\hat{H}^{y}_{0}(\eta)$ and $\hat{H}^{\varphi}_{0}(\eta)$ are the free Hamiltonian of the tensor perturbation $\hat{y}_{ij}$ and the other fields $\hat{\varphi}$, respectively. The operator $\hat{V}(\eta)$ is the interaction between the tensor perturbation and the other fields. We assume that the initial state of the total system $\ket{\Psi}$ at $\eta \rightarrow -\infty$ is given by the product state 
\begin{equation}
\ket{\Psi}=\ket{0^\text{BD}_{y}} \otimes \ket{\psi_\varphi}, \label{eq:Psi}
\end{equation}
where $\ket{0^\text{BD}_{y}}$ is the Bunch-Davies vacuum of the tensor field and $\ket{\psi_\varphi}$ is the initial state of the other fields. The wave functional of the total system is 
\begin{equation}
\Psi_{\eta}[y, \varphi]=\bra{y,\varphi} \hat{U}(\eta, -\infty) \ket{\Psi}, \label{eq:Psieta}
\end{equation}
where the time evolution operator $\hat{U}(\eta,-\infty)$ is expressed
by using the time ordering as
\begin{equation}
\hat{U}(\eta,-\infty)=\text{T}\exp \left[-i \int^{\eta}_{-\infty} d\tau \hat{H}(\tau) \right].  \label{eq:U}
\end{equation}
We give the decohered state by assuming the following form of the reduced density matrix of $y_\lambda$ : 
\begin{equation}
\rho_{\eta} [y, y']= \int_{\varphi} \Psi_{\eta}[y, \varphi] \Psi^{*}_{\eta}[y', \varphi] =\Psi^\text{BD}_{\eta}[y] (\Psi^\text{BD}_{\eta}[y'])^{*} D_{\eta}[y,y'], \label{eq:decstate}
\end{equation}
with $\Psi^\text{BD}_{\eta}[y]$ and $D_{\eta}[y,y']$ are 
\begin{align}
\Psi^\text{BD}_{\eta}[y]&=N(\eta) \exp \left[-\frac{1}{2} \sum_{\lambda=1,2} \int d^{3}q \, \Omega_q (\eta)|y_\lambda (\bm{q})|^{2} \right], \label{eq:PsiBD} \\
D_{\eta}[y,y']&= \exp \left[-\frac{1}{2} \sum_{\lambda=1,2} \int d^{3}q \, \Gamma_q (\eta)|y_\lambda (\bm{q})-y'_\lambda (\bm{q})|^{2}  \right], \label{eq:Dyy}
\end{align}
where $N(\eta)$ is the normalization and $\Omega_q (\eta)$ is given by
\eqref{eq:psiBD}. A phenomenological positive function
  $\Gamma_q(\eta)$ represents decoherence effect. The function
$\Gamma_q$ depends on the structure of interaction with other fields
$\varphi$. The decoherence factor $D_\eta [y,y']$ is invariant under
spatial rotations and translations, which preserves the same symmetry
imposed in the linear theory of PGWs. As $\Gamma_q$ becomes large, the
off-diagonal components $\rho_\eta [y,y']$ decays exponentially. This
behavior expresses  quantum decoherence.  The form of the
decoherence factor $D_\eta[y,y']$ respects the facts that the field
operator (growing mode) is the pointer observable \cite{Zurek1981} in
cosmology. For super-horizon modes, in the Heisenberg picture, the
field becomes constant in time and its conjugate momentum decays
rapidly. Thus the field operator effectively commutes with the
interaction Hamiltonian. Such an operator
commuting with the interaction Hamiltonian is called a pointer
observable. The density matrix of the system approaches diagonal form
with respect to the basis of the pointer observable (pointer basis) by
 decoherence effect.
In \cite{Kiefer2007, Burgess2008, Burgess2015}, for the super-horizon
mode ($q\eta \ll 1$), the decoherence factor was derived using the
quantum master equations with the Lindblad form \cite{Lindblad1976,
  Breuer2007}. Also the decoherence factor were computed from
nonlinear interactions for primordial fluctuations in
\cite{Nelson2016, Boyanovsky2016, Hollowood2017}.

In \cite{Kiefer2007}, the authors
focused on the Wigner function of the density matrix of the decohered
state and discussed its shape in the phase space. The density matrix
$\rho(y,y',\eta)$ for a fixed mode $\bm{q}$ and polarization $\lambda$
is
\begin{equation}
\rho(y,y', \eta)=\psi^\text{BD}(y,\eta)[\psi^{\text{BD}}(y',\eta)]^{*} \exp \left(-\Gamma_q|y-y'|^{2} \right),  \label{eq:decstate2} 
\end{equation}
where $\psi^\text{BD} (y,\eta)$ is the wave function of the
Bunch-Davies given in \eqref{eq:psiBD}. The real part
$\Omega^\text{R}_q$ characterizes the quantum superposition
with respect to the field basis $y$. Such a superposition is
suppressed by the decoherence factor if the parameter $\Gamma_q$
satisfies the inequality
\begin{equation}
\Gamma_q \gg \Omega_q^\text{R}. \qquad(\text{decoherence condition}) \label{eq:decocond}
\end{equation}
The decoherence degrades the superposition of the field amplitudes and
makes the width of the Wigner function large in the direction of the
conjugate momentum as follows. The Wigner function of the density
matrix $\rho(y,y',\eta)$ is
\begin{align}
W(y,\pi_{y}, \eta)&=w(y^\text{R}, \pi^\text{R}_{y},\eta) w(y^\text{I}, \pi^\text{I}_{y},\eta), \label{eq:Wigner2} \\
w(x, p,\eta)&=\sqrt{ \frac{\Omega_q^\text{R}}{\pi^2 (\Omega_q^\text{R} + 2\Gamma_q)}} \exp \left[ -2\Omega_q^\text{R}\, x^{2} -2\frac{\left(p +\Omega_{q}^\text{I} \, x\right)^{2}}{\Omega_q^\text{R} +2\Gamma_q} \right]. \label{eq:wigner}
\end{align}
  For a large $\Gamma_q$, the Gaussian width for the conjugate
  momentum becomes large, and then Wigner ellipse approaches a
  circle. To observe the oscillation of the angular-power spectrum,
  the Wigner function should be squeezed even if decoherence
  occurs. In terms of the length of the major axis $a$ and the
minor axis $b$ of the Wigner ellipse, the condition of squeezing \cite{Kiefer2007} is
expressed as 
\begin{equation}
a \gg b.  \qquad (\text{correlation condition})\label{eq:correcond}
\end{equation}
The word ``correlation'' does not mean quantum correlations but the
correlation between the real (or imaginary) part of the field variable
and its conjugate momentum.

\begin{figure}[H]
   \centering
   \includegraphics[width=0.50\linewidth]{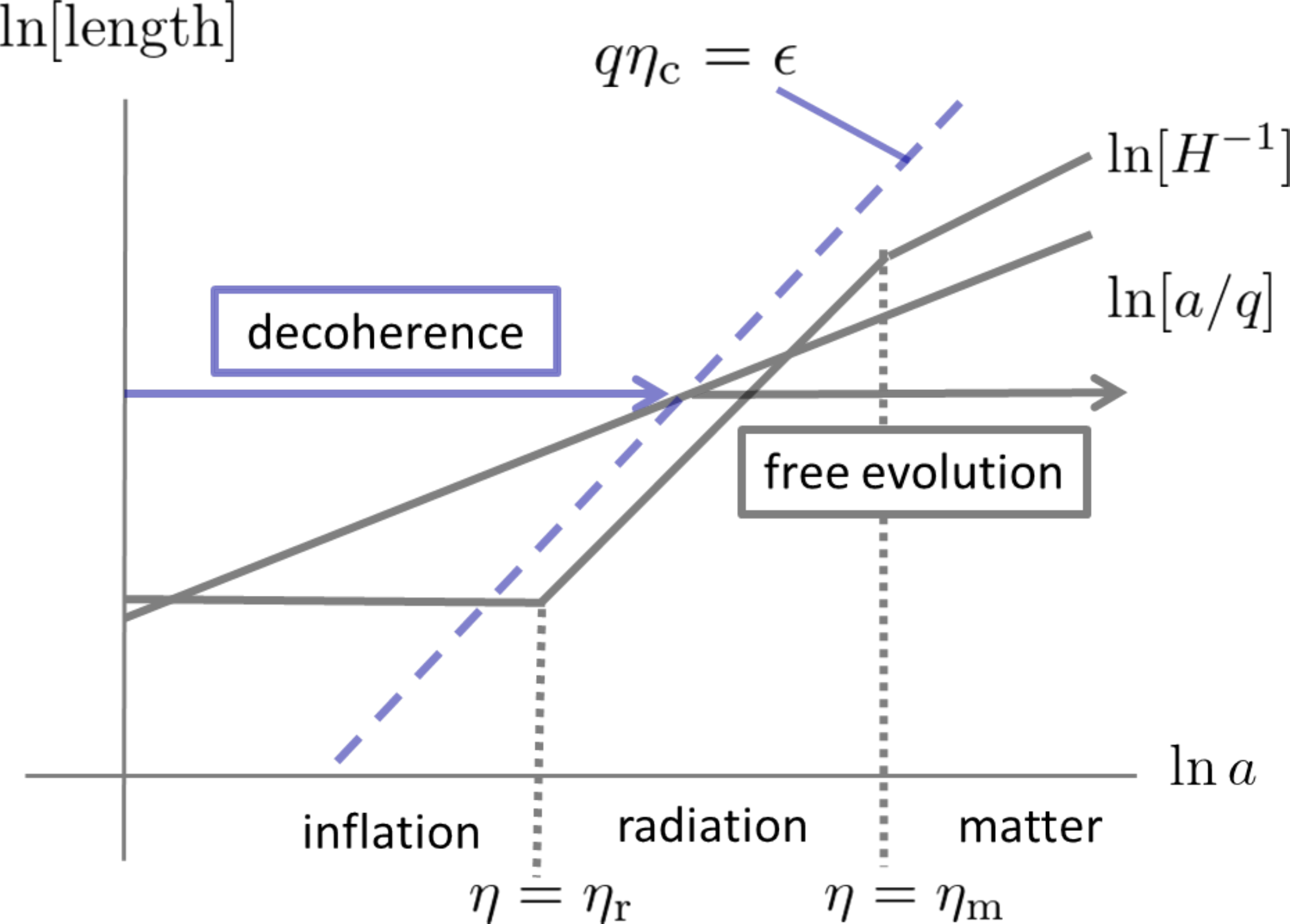}
 \caption{The assumption for the evolution of PGWs. Quantum decoherence continues until $\eta_\text{c}$ and then the PGW evolves unitarily. }
    \label{fig:deco}
\end{figure}

In the following, we clarify the relation among the quantum
correlations of PGWs at the matter era and the above conditions
\eqref{eq:decocond} and \eqref{eq:correcond}. For this purpose we
consider the scenario that the decoherence due to the interaction
halts just before the second horizon crossing of PGWs and the state of
PGWs evolves unitarily after that. In this scenario, the decohered
state of PGWs \eqref{eq:decstate2} is prepared at the conformal time
$\eta_\text{c}$ which satisfies
\begin{equation}
\eta_\text{r} \leq \eta_\text{c}, \quad q\,\eta_\text{c}  =\epsilon, \label{eq:etac}
\end{equation}
where $\epsilon \sim 1$ is a model parameter. The whole evolution of PGWs in our setting is presented in Fig.~\ref{fig:deco}.
We examine the decoherence condition \eqref{eq:decocond} and
the correlation condition \eqref{eq:correcond} at
$\eta=\eta_\text{c}$. To observe the decohered but squeezed
  state of PGWs, these conditions should be satisfied at the horizon
  crossing $\epsilon \sim1$. For a super-horizon mode at
    $\eta_r$, $q\eta_\text{r} \ll 1$, the decoherence condition 
is estimated as 
\begin{equation}
\frac{1}{|\beta_q|^{2}}  \ll \frac{\Gamma_q (\eta_\text{c})}{q}\label{eq:decocond2}
\end{equation}
and the correlation condition is given as 
\begin{equation}
\frac{\Gamma_q (\eta_\text{c})}{q} \ll |\beta_q|^{2},  \label{eq:correcond2}
\end{equation}
where $\beta_q$ is the Bogolyubov coefficient given in \eqref{eq:alphabeta}. 

Let us investigate the entanglement and quantum discord of PGWs in the matter era. For $\eta,\eta'>\eta_c$, we
have the two-point function $\langle yy \rangle$
\begin{equation}
\bra{\Psi} \hat{y}^\text{H}_\lambda (\bm{q},\eta) \hat{y}^\text{H}_{\lambda'} (\bm{q}',\eta') \ket{\Psi}
=\bra{\Psi}\hat{\Omega}^{\dagger}(\eta_\text{c},-\infty) \hat{y}^\text{I}_\lambda (\bm{q},\eta) \hat{y}^\text{I}_{\lambda'}(\bm{q}',\eta') \hat{\Omega}(\eta_\text{c},-\infty) \ket{\Psi}, \label{eq:yHyH} 
\end{equation}
where $\hat{y}^\text{H}_\lambda$ and $\hat{y}^\text{I}_\lambda$ are
the tensor field in the Heisenberg and interaction picture,
respectively and $\hat{\Omega}(\eta,-\infty)$ is given by
\begin{equation}
\hat{\Omega}(\eta,-\infty)=\text{T}\exp \left[-i \int^{\eta}_{-\infty} d\tau \hat{V}^\text{I}(\tau) \right]. \label{eq:Omega}
\end{equation}
The concrete expression of the interaction Hamiltonian is not needed
because the reduced density matrix of the tensor field
\eqref{eq:decstate} is given at $\eta_\text{c}$.  In Eq.~\eqref{eq:yHyH}, 
we assumed that the interaction continues until
$\eta_\text{c}$, that is,
$\hat{\Omega}(\eta,-\infty)=\hat{\Omega}(\eta_\text{c},-\infty)$ for
$\eta_c \leq \eta$. The field operator $\hat{y}^\text{I}_\lambda (\bm{q}, \eta)$ can be written by the linear combination of $\hat{y}^\text{I}_\lambda (\bm{q}, \eta_\text{c})$ and $\hat{\pi}^\text{I}_\lambda (\bm{q}, \eta_\text{c})$ at $\eta_\text{c}$ as
\begin{equation}
\hat{y}^\text{I}_\lambda(\bm{q},\eta)
=X_q (\eta,\eta_\text{c}) \hat{y}^\text{I}_\lambda(\bm{q},\eta_\text{c})+Y_{q}(\eta,\eta_\text{c}) \hat{\pi}^\text{I}_\lambda (\bm{q},\eta_\text{c}), \label{eq:yforycpic} 
\end{equation}
where $X_q$ and $Y_q$ are defined by
\begin{align}
X_q (\eta,\eta')&=f_q (\eta)g^{*}_q (\eta')+f^{*}_q (\eta)g_q (\eta'), \label{eq:Xq} \\
Y_q (\eta,\eta')&=i[f_q (\eta)f^{*}_q (\eta')-f^{*}_q (\eta)f_q (\eta')]. \label{eq:Yq} 
\end{align}
From the form of the density matrix at $\eta_\text{c}$
\eqref{eq:decstate}, the correlation functions of the tensor field at the time $\eta_\text{c}$ in the interaction picture can be computed as follows:
\begin{align}
\bra{\Psi}\hat{\Omega}^{\dagger}(\eta_\text{c},-\infty) \hat{y}^\text{I}_\lambda (\bm{q},\eta_\text{c}) \hat{y}^\text{I}_{\lambda'} (\bm{q}',\eta_\text{c}) \hat{\Omega}(\eta_\text{c},-\infty) \ket{\Psi} 
&= \bra{0^\text{BD}_y} \hat{y}^\text{I}_\lambda (\bm{q},\eta_\text{c}) \hat{y}^\text{I}_{\lambda'} (\bm{q}',\eta_\text{c}) \ket{0^\text{BD}_y}, \label{eq:ycyc}  \\
\bra{\Psi}\hat{\Omega}^{\dagger}(\eta_\text{c},-\infty) \hat{y}^\text{I}_\lambda (\bm{q},\eta_\text{c}) \hat{\pi}^\text{I}_{\lambda'} (\bm{q}',\eta_\text{c}) \hat{\Omega}(\eta_\text{c},-\infty) \ket{\Psi} &= \bra{0^\text{BD}_y} \hat{y}^\text{I}_\lambda (\bm{q},\eta_\text{c}) \hat{\pi}^\text{I}_{\lambda'}(\bm{q}',\eta_\text{c}) \ket{0^\text{BD}_y}, \label{eq:ycpic} \\
\bra{\Psi}\hat{\Omega}^{\dagger}(\eta_\text{c},-\infty) \hat{\pi}^\text{I}_\lambda (\bm{q},\eta_\text{c}) \hat{\pi}^\text{I}_{\lambda'}(\bm{q}',\eta_\text{c}) \hat{\Omega}(\eta_\text{c},-\infty) \ket{\Psi} 
&= \bra{0^\text{BD}_y} \hat{\pi}^\text{I}_\lambda (\bm{q},\eta_\text{c}) \hat{\pi}^\text{I}_{\lambda'} (\bm{q}',\eta_\text{c}) \ket{0^\text{BD}_y} \nonumber \\
&+\Gamma_q (\eta_\text{c}) \delta_{\lambda \lambda'} \delta^{3}(\bm{q}+\bm{q}'). \label{eq:picpic} 
\end{align}
The derivation of these equations is presented in the appendix
C. Substituting Eq.~\eqref{eq:yforycpic} into the correlator
\eqref{eq:yHyH} and using the formulas \eqref{eq:ycyc},
\eqref{eq:ycpic} and \eqref{eq:picpic}, we obtain the correlator
\eqref{eq:yHyH} for the different time $\eta$ and $\eta'$ as
\begin{align}
\bra{\Psi} \hat{y}^\text{H}_\lambda (\bm{q},\eta) \hat{y}^\text{H}_{\lambda'} (\bm{q}',\eta') \ket{\Psi}
&=\bra{0^\text{BD}_y} \hat{y}^\text{I}_\lambda (\bm{q},\eta) \hat{y}^\text{I}_{\lambda'} (\bm{q}',\eta') \ket{0^\text{BD}_y} \nonumber \\
&+Y_{q}(\eta,\eta_\text{c})Y_{q}(\eta',\eta_\text{c})\Gamma_{q}(\eta_\text{c})\delta_{\lambda \lambda'} \delta^{3}(\bm{q}+\bm{q}').  \label{eq:yydec} 
\end{align}
We can also calculate the other
two-point functions $\langle y\pi \rangle$ and
$\langle \pi \pi \rangle$. The conjugate momentum $\hat{\pi}^\text{I}_\lambda (\bm{q},\eta)$ is given by the following linear combination of $\hat{y}^\text{I}_\lambda (\bm{q},\eta_\text{c})$ and $\hat{\pi}^\text{I}_\lambda (\bm{q}, \eta_\text{c})$:
\begin{equation}
\hat{\pi}^\text{I}_\lambda(\bm{q},\eta)
=z_q (\eta,\eta_\text{c})
\hat{y}^\text{I}_\lambda(\bm{q},\eta_\text{c})+w_{q}(\eta,\eta_\text{c})
\hat{\pi}^\text{I}_\lambda
(\bm{q},\eta_\text{c}), \label{eq:piforycpic}  
\end{equation}
where $z_q$ and $w_q$ are defined by
\begin{align}
z_q (\eta,\eta')&=(-i)\left[g_q (\eta)g^{*}_q (\eta')-g^{*}_q (\eta)g_q (\eta')\right], \label{eq:Zq} \\
w_q (\eta,\eta')&=g_q (\eta)f^{*}_q (\eta')+g^{*}_q (\eta)f_q (\eta'). \label{eq:Wq}
\end{align}
Through the similar procedure, we can derive the other correlators as
\begin{align}
\bra{\Psi} \hat{y}^\text{H}_\lambda (\bm{q},\eta) \hat{\pi}^\text{H}_{\lambda'} (\bm{q}',\eta') \ket{\Psi}
&=\bra{0^\text{BD}_y} \hat{y}^\text{I}_\lambda (\bm{q},\eta) \hat{\pi}^\text{I}_{\lambda'} (\bm{q}',\eta') \ket{0^\text{BD}_y} \nonumber \\
&+Y_{q}(\eta,\eta_\text{c})w_{q}(\eta',\eta_\text{c})\Gamma_{q}(\eta_\text{c})\delta_{\lambda \lambda'}\delta^{3}(\bm{q}+\bm{q}'),  \label{eq:ypdec} \\
\bra{\Psi} \hat{\pi}^\text{H}_\lambda (\bm{q},\eta) \hat{\pi}^\text{H}_{\lambda'} (\bm{q}',\eta') \ket{\Psi}
&=\bra{0^\text{BD}_y} \hat{\pi}^\text{I}_\lambda (\bm{q},\eta) \hat{\pi}^\text{I}_{\lambda'} (\bm{q}',\eta') \ket{0^\text{BD}_y} \nonumber \\
&+w_{q}(\eta,\eta_\text{c})w_{q}(\eta',\eta_\text{c})\Gamma_{q}(\eta_\text{c})\delta_{\lambda \lambda'}\delta^{3}(\bm{q}+\bm{q}').  \label{eq:ppdec}
\end{align}
By Eqs.~\eqref{eq:yydec},
\eqref{eq:ypdec} and \eqref{eq:ppdec}, the correlators of $\hat{A}^\text{H}_\lambda$  and $\hat{A}^{\text{H}\dagger}_\lambda$ at  $\eta$ are given by
\begin{align}
\bra{\Psi} \hat{A}^{\text{H} \dagger}_\lambda (\bm{q},\eta)\hat{A}^\text{H}_{\lambda'} (\bm{q}',\eta) \ket{\Psi}
&=n^\text{dec}_q (\eta) \delta_{\lambda \lambda'} \delta^{3}(\bm{q}-\bm{q}'), \label{eq:AdgAdec} \\
\bra{\Psi} \hat{A}^\text{H}_\lambda (\bm{q},\eta)\hat{A}^\text{H}_{\lambda'} (\bm{q}',\eta) \ket{\Psi}
&=c^\text{dec}_q (\eta) \delta_{\lambda \lambda'} \delta^{3}(\bm{q}+\bm{q}'), \label{eq:AAdec} 
\end{align}
where we introduced the following quantities
\begin{align}
n^\text{dec}_{q}(\eta)&:=n_q (\eta) +\left| \sqrt{\frac{q}{2}}\,Y_{q}(\eta,\eta_\text{c})+\frac{i}{\sqrt{2q}} w_{q}(\eta,\eta_\text{c}) \right|^{2} \Gamma_q (\eta_\text{c}), \label{eq:Ndec} \\
c^\text{dec}_{q}(\eta)
&:=c_q (\eta)+\left( \sqrt{\frac{q}{2}}\,Y_{q}(\eta,\eta_\text{c})+\frac{i}{\sqrt{2q}} w_{q}(\eta,\eta_\text{c}) \right)^{2} \Gamma_q (\eta_\text{c}). \label{eq:Cdec} 
\end{align}
We focus on the target wave mode
$1/\eta_\text{m} \ll q \ll 1/\eta_\text{r}$ \eqref{eq:qobs} and
 examine the PPT criterion in the matter era
$\eta>\eta_\text{m}$. The decohered state is the bipartite state with
the mode $\bm{q}$ and $-\bm{q}$ defined by the annihilation operators $\hat{A}^\text{H}(\bm{q}, \eta)$ and
$\hat{A}^{\text{H}}(-\bm{q},\eta)$. For the sub-horizon mode, the
operator $\hat{A}^\text{H}(\bm{q}, \eta)$ is the counterpart of $\hat{b}_\lambda (\bm{q})$ due to
the relation $\hat{A}^\text{I} \sim \hat{b}\exp(-iq\eta)$
(Eq.~\eqref{eq:approxA}). Using Eqs~\eqref{eq:Ndec} and
\eqref{eq:Cdec}, we can rewrite the PPT criterion \eqref{eq:PPT2}
$n^\text{dec}_q\geq|c^\text{dec}_q|$ as
\begin{equation}
 \frac{\Gamma_q (\eta_\text{c})}{q} \geq \frac{n_q (\eta)}{|qY_q
   (\eta,\eta_\text{c})+iw_q (\eta,\eta_\text{c})|^{2}\, n_q
   (\eta_\text{c})-\text{Re}[c_q (\eta)(qY_{q}(\eta,\eta_\text{c})-iw_q (\eta,\eta_\text{c}))^{2}]}. \label{eq:PPTdec} 
\end{equation}
For $q\eta_{c}=\epsilon \sim 1$, this inequality is evaluated up to the numerical factor as 
\begin{equation}
 \frac{\Gamma_q (\eta_\text{c})}{q} \gtrsim 1,
\label{eq:PPTdec2} 
\end{equation}
where we used the approximated
formulas \eqref{eq:approxnq}, \eqref{eq:approxcq} and
\begin{equation}
qY_q (\eta,\eta_\text{c})+iw_q (\eta,\eta_\text{c}) \sim i e^{-iq(\eta -\eta_\text{c})} \label{eq:approxqYW} 
\end{equation}
for a sub-horizon scale $q\eta \gg 1$. 

For the target frequency $q \eta_\text{r} \ll 1$, the tensor fields have the large
occupation number $|\beta_q|^{2} \gg 1$, and the PPT criterion
\eqref{eq:PPTdec2} implies the decoherence condition
\eqref{eq:decocond2}
\begin{equation}
 \frac{\Gamma_q (\eta_\text{c})}{q} \gtrsim 1 \quad \Longrightarrow \quad \frac{\Gamma_q(\eta_\text{c})}{q} \gg \frac{1}{|\beta_q|^{2}}
\label{eq:PPTtodeco} 
\end{equation}
Hence the decoherence condition \eqref{eq:decocond2} is not sufficient
to eliminate the entanglement of PGWs. Next we evaluate the degree of
quantum coherence $c^\text{dec}_q (\eta)$ to examine the quantum
discord of PGWs.  For the target wave number
$1/\eta_\text{m} \ll q \ll 1/\eta_\text{r}$, we can approximate the
function $c^\text{dec}_q (\eta)$ as
\begin{equation}
c^\text{dec}_{q}(\eta) \sim  -\left( |\beta_q|^{2}  +\frac{\Gamma_q (\eta_\text{c})}{2q} e^{2i \epsilon} \right)e^{-2iq\eta}, \label{eq:approxCdec}
\end{equation}
where we applied the approximated formulas \eqref{eq:approxnq},
\eqref{eq:approxcq} and \eqref{eq:approxqYW} again. If the
phenomenological parameter $\Gamma_q(\eta_c)$ satisfies the correlation
condition \eqref{eq:correcond2}, then the decoherence effect
is negligible in \eqref{eq:approxCdec}. In this case, the
quantum coherence of the Bunch-Davies vacuum survives. Because the
decohered state is a Gaussian state, the nonzero $c^\text{dec}_q$ implies 
quantum discord in the matter-dominated era. 
Hence the correlation condition given in
\cite{Kiefer2007} means that the PGWs have a chance to keep the
quantum discord in the matter-dominated era.

\begin{figure}[H]
   \centering
   \includegraphics[width=0.495\linewidth]{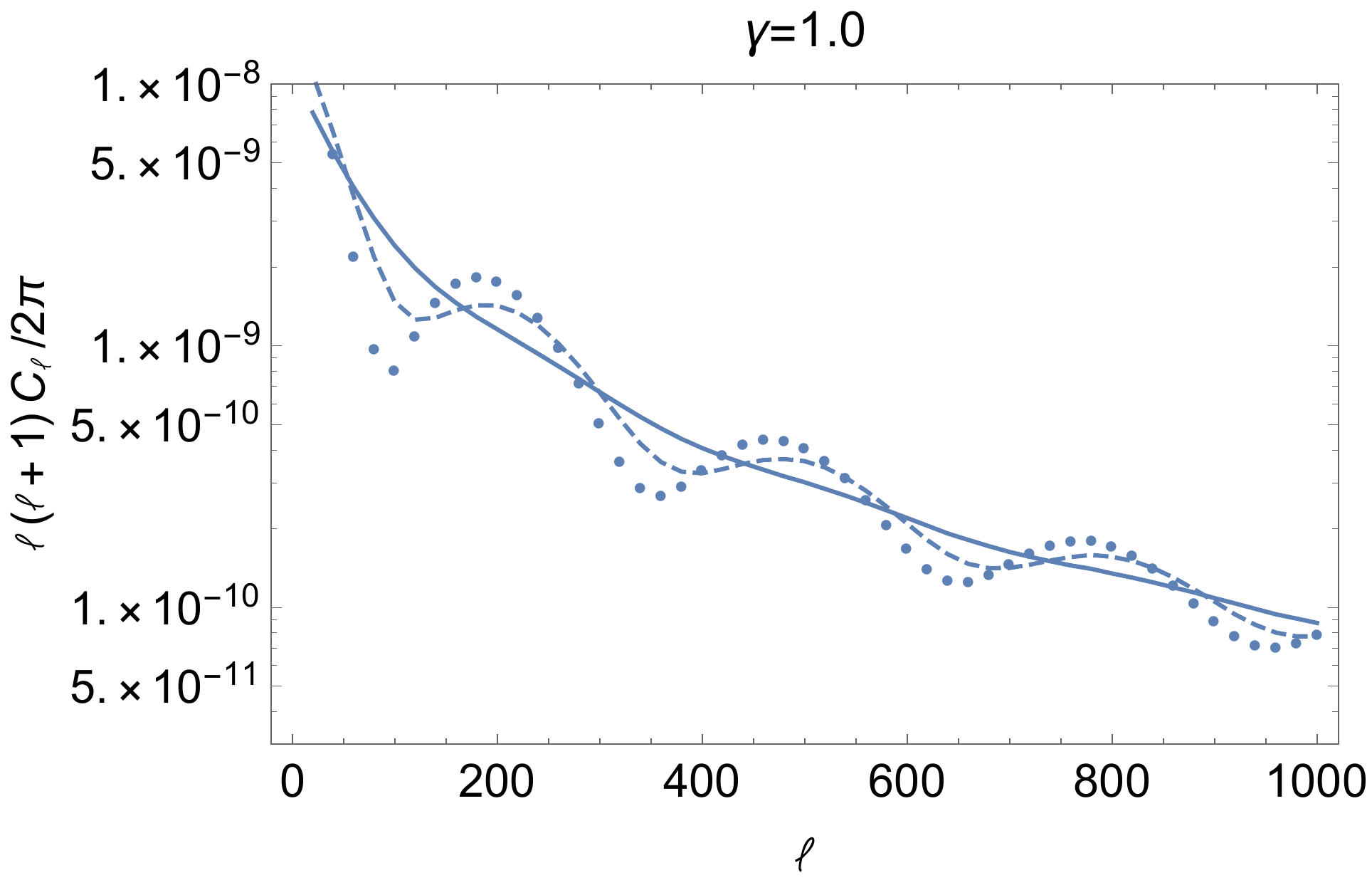}
   \includegraphics[width=0.495\linewidth]{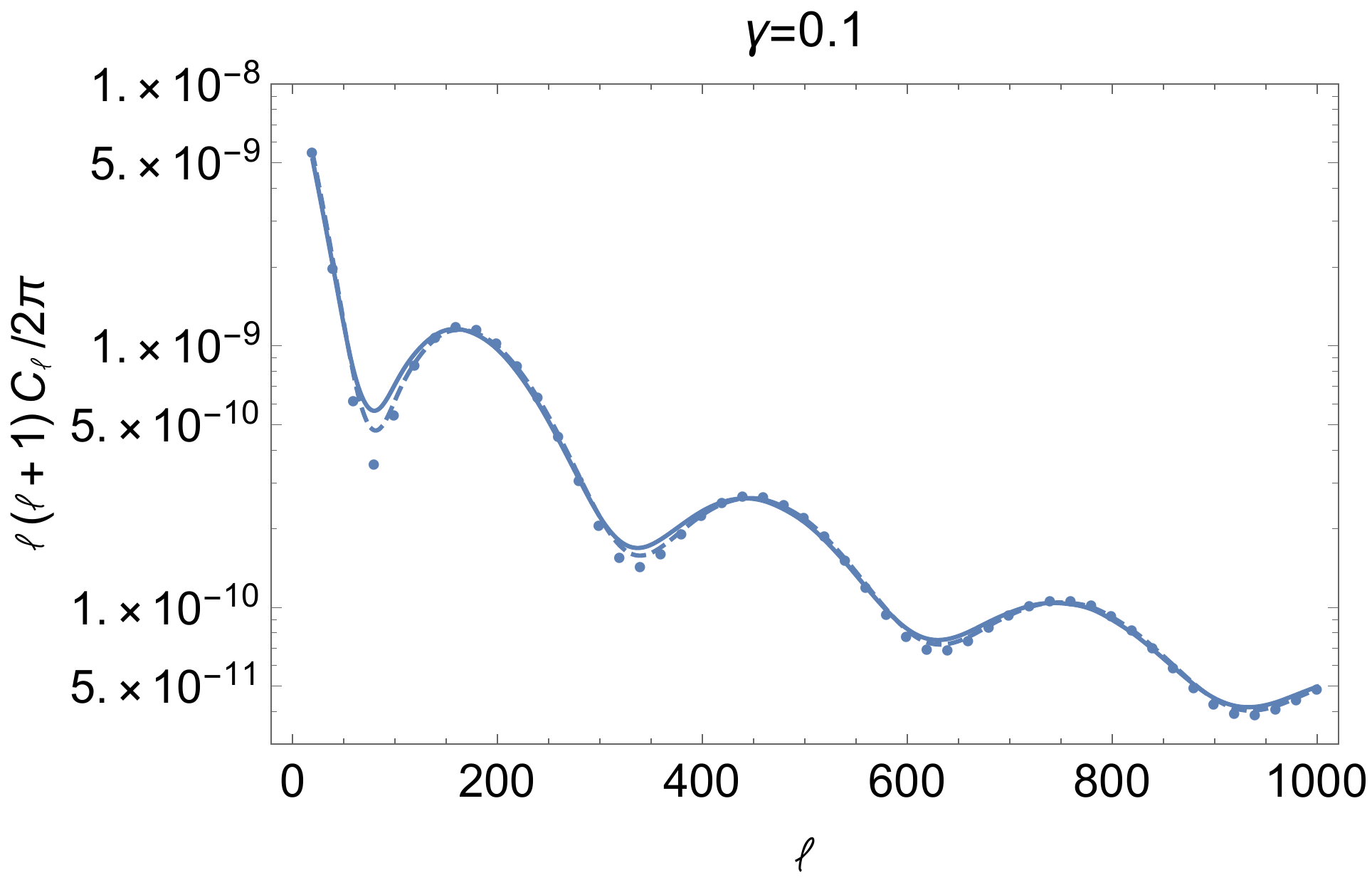}
   \caption{The anguler power spectrum of CMB fluctuations by PGWs
     with the decoherence effect (left panel: $\gamma=1.0$ and right panel:
     $\gamma=0.1$). The different curves correspond to $\epsilon=0.5$
     (dotted line), $\epsilon=1.0$ (dashed line) and $\epsilon=1.5$
     (solid line).}
    \label{fig:largesmall}
\end{figure}

\begin{figure}[H]
   \centering
   \includegraphics[width=0.70\linewidth]{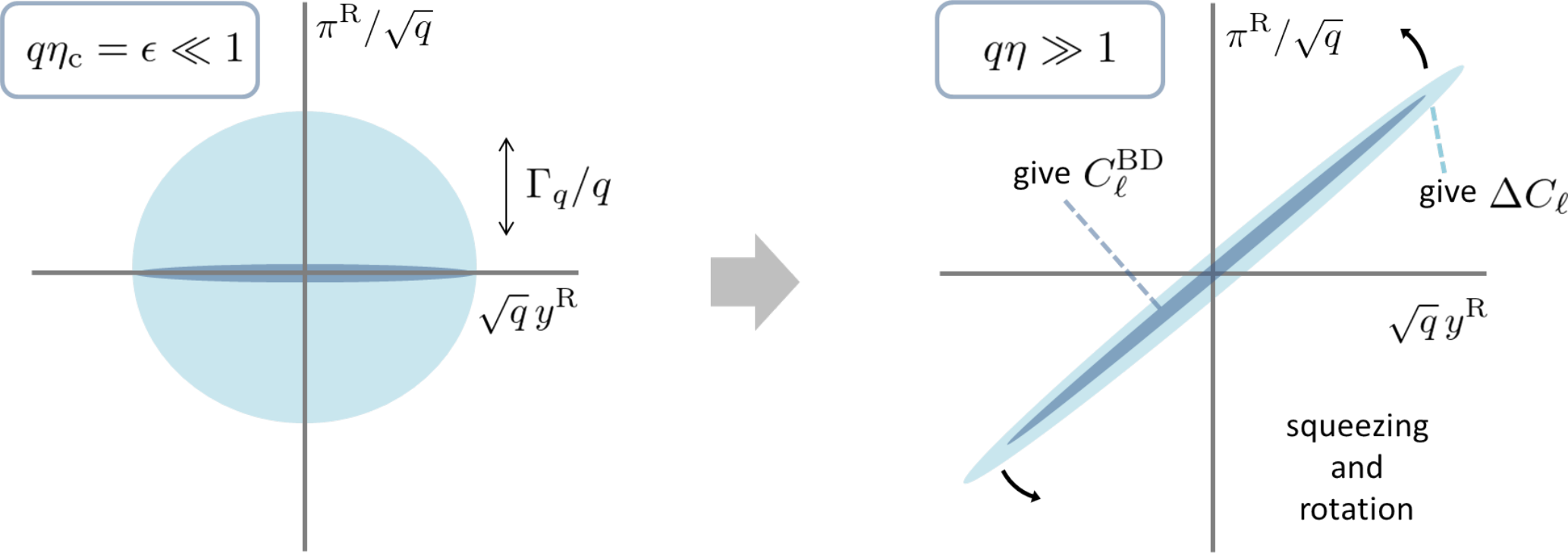}
   \caption{For the super-horizon mode, the Wigner function is
     squeezed until the mode re-enters the horizon after 
       decoherence.}
    \label{fig:ypideco}
\end{figure}

Let us demonstrate the behavior of the angular-power spectrum for
the decohered state.  By the formula \eqref{eq:ppdec}, the
angular-power spectrum $C^\text{dec}_\ell$ for the decohered state is
given by
\begin{equation}
C^\text{dec}_\ell=C^\text{BD}_{\ell}+\Delta C_\ell, \label{eq:Cdecl}
\end{equation}
where the impact of the decoherence on the
angular-power spectrum is represented as
\begin{equation}
\Delta C_\ell=\frac{8\pi}{2\ell+1} \int^{\infty}_{0} dq q^{2} \, \Gamma_q (\eta_\text{c})|W_{\ell} (q)|^{2}, \label{eq:DeltaC}
\end{equation}
with 
\begin{equation}
W_\ell (q):=\frac{\sqrt{2}}{M_\text{pl}} \frac{1}{(2\pi)^{3/2}} \sqrt{\frac{\pi(2\ell+1)(\ell+2)!}{2(\ell-2)!}} \int^{\eta_0}_{\eta_\text{L}} \frac{d\eta}{a(\eta)} \frac{j_{\ell}(q(\eta_0 -\eta))}{q^{2}(\eta_0 - \eta)^{2}} w_{q} (\eta, \eta_\text{c}). \label{eq:Wl} 
\end{equation}
In principle, the function $\Gamma_q (\eta)$ can be
determined by assuming nonlinear interactions with
other fields. Since a macroscopic system easily decoheres, we can expect that the value of $\Gamma_q (\eta_\text{c})$ increases for the larger system. For
simplicity we assume that $\Gamma_q (\eta_\text{c})$ per mode is
proportional to the number density $|\beta_q|^{2}$, that is,
\begin{equation}
\frac{\Gamma_{q}(\eta_\text{c})}{q}=\gamma |\beta_q|^{2}, \label{eq:Gamma} 
\end{equation}
where $\gamma$ is a dimensionless positive constant. For
$\gamma \sim 1$, the correlation condition \eqref{eq:correcond2} is
violated. In Fig.~\ref{fig:largesmall}, we present the behavior of
$\ell(\ell+1)C^\text{dec}_{\ell}/2\pi$ for $\gamma=1.0$ and
$\gamma=0.1$ with $\epsilon=0.5, 1.0, 1.5$.  As have already
mentioned, the decoherence changes the ellipse of the Wigner function
to a circle and hence the observable oscillation is reduced. However, in the left panel of Fig.~\ref{fig:largesmall} for $\gamma=1.0$,  we still
observe the oscillation after the decoherence for the super-horizon
mode $\epsilon =0.5$ even if the correlation condition
\eqref{eq:correcond2} is violated.  This is because
the Wigner function of PGWs with the super-horizon mode is squeezed
until the horizon crossing after the decoherence (see
Fig.~\ref{fig:ypideco}). 

\noindent
We observe that the oscillation vanishes for $\epsilon=1.5$. In
this case, the Wigner ellipse becomes a circle and its shape does not
change after the decoherence because of no squeezing effect for the
sub-horizon modes. In the right panel of Fig.~\ref{fig:largesmall}, 
we show the behavior of $\ell(\ell+1)C^\text{dec}_{\ell}/2\pi$ for
 $\gamma=0.1$. The oscillation does not vanish since the quantum discord of PGWs survives for $\gamma=0.1$ (in other words, the correlation 
condition is satisfied). 

\begin{figure}[H]
   \centering
   \includegraphics[width=0.495\linewidth]{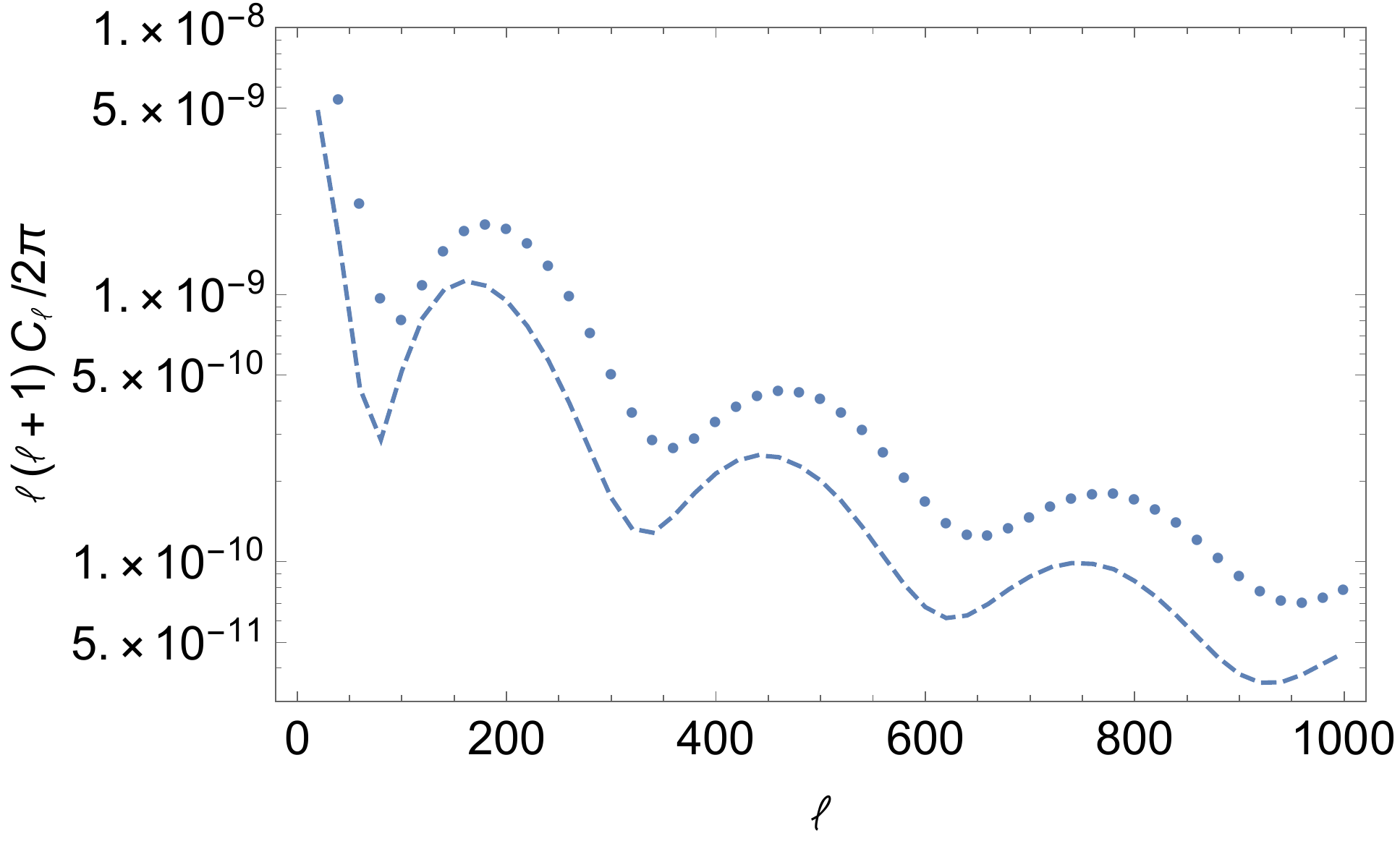}
   \includegraphics[width=0.495\linewidth]{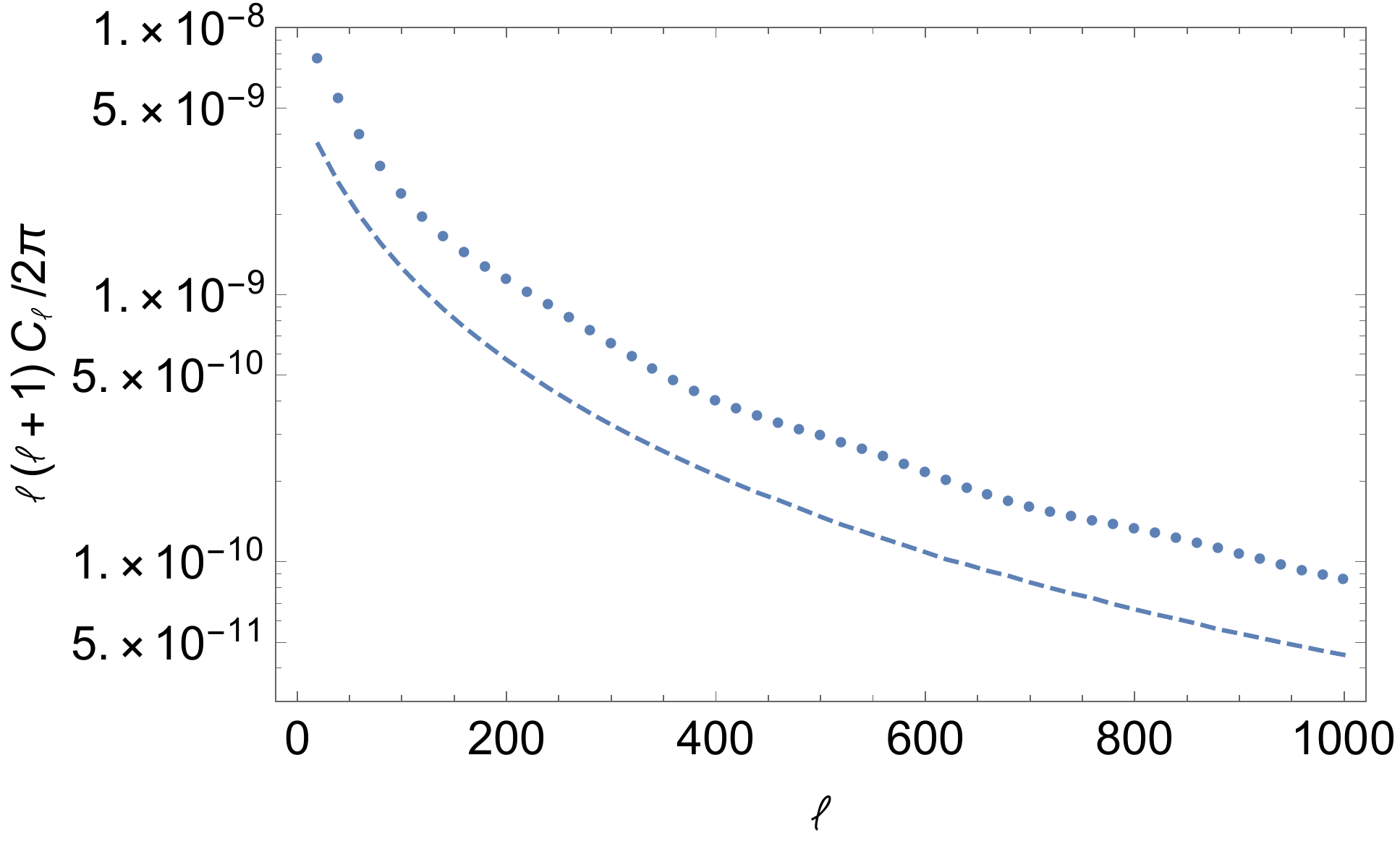}
   \caption{Left panel: $C^\text{BD}_\ell$ (dashed line) and
     $C^\text{dec}_\ell$ (dotted line)  with $\epsilon=0.5$. Right panel: 
     $C^\text{cl}_\ell$ (dashed line) and $C^\text{dec}_\ell$
     (dotted line) with $\epsilon=1.5$. }
    \label{fig:BDclvsdec}
\end{figure}
In Fig.~\ref{fig:BDclvsdec}, we compare the angular-power spectrum for
the Bunch-Davies vacuum and the classical state with the
decoherence effect ($\gamma=1.0$). The left panel presents the behaviors of $C^\text{BD}_\ell$ and
$C^\text{dec}_\ell$ with $\epsilon=0.5$ which show oscillation. The
right panel shows the behaviors of
$C^\text{cl}_\ell$ and $C^\text{dec}_\ell$ with
$\epsilon=1.5$. The oscillations are reduced by the decoherence effect.
In this case, $C^\text{dec}_\ell$ is almost $2C^\text{cl}_\ell$. For
$\gamma=1.0$ and $\epsilon=1.5$, $\Delta C_\ell$ has the same
amplitude and almost opposite phase as $C^\text{BD}_\ell$. That
is $\Delta C_\ell$ can be evaluated by $C^\text{BD}_\ell$
using the mode function $e^{i\pi/2} v^\text{rad}_q$. Thus we find
that
\begin{equation}
\Delta C_\ell \sim \frac{16\pi}{2\ell+1} \int^{\infty}_{0} dq\, q^{2} \, |\beta_q|^2 \Bigl[ |V_\ell (q)|^{2}- V^{2}_{\ell}(q)/2- V^{*2}_{\ell}(q)/2 \Bigr],
\end{equation}
and $C^\text{dec}_\ell=C^\text{BD}_\ell +\Delta C_\ell \sim 2 C^\text{cl}_\ell$.

\begin{figure}[H]
  \centering
   \includegraphics[width=0.50\linewidth]{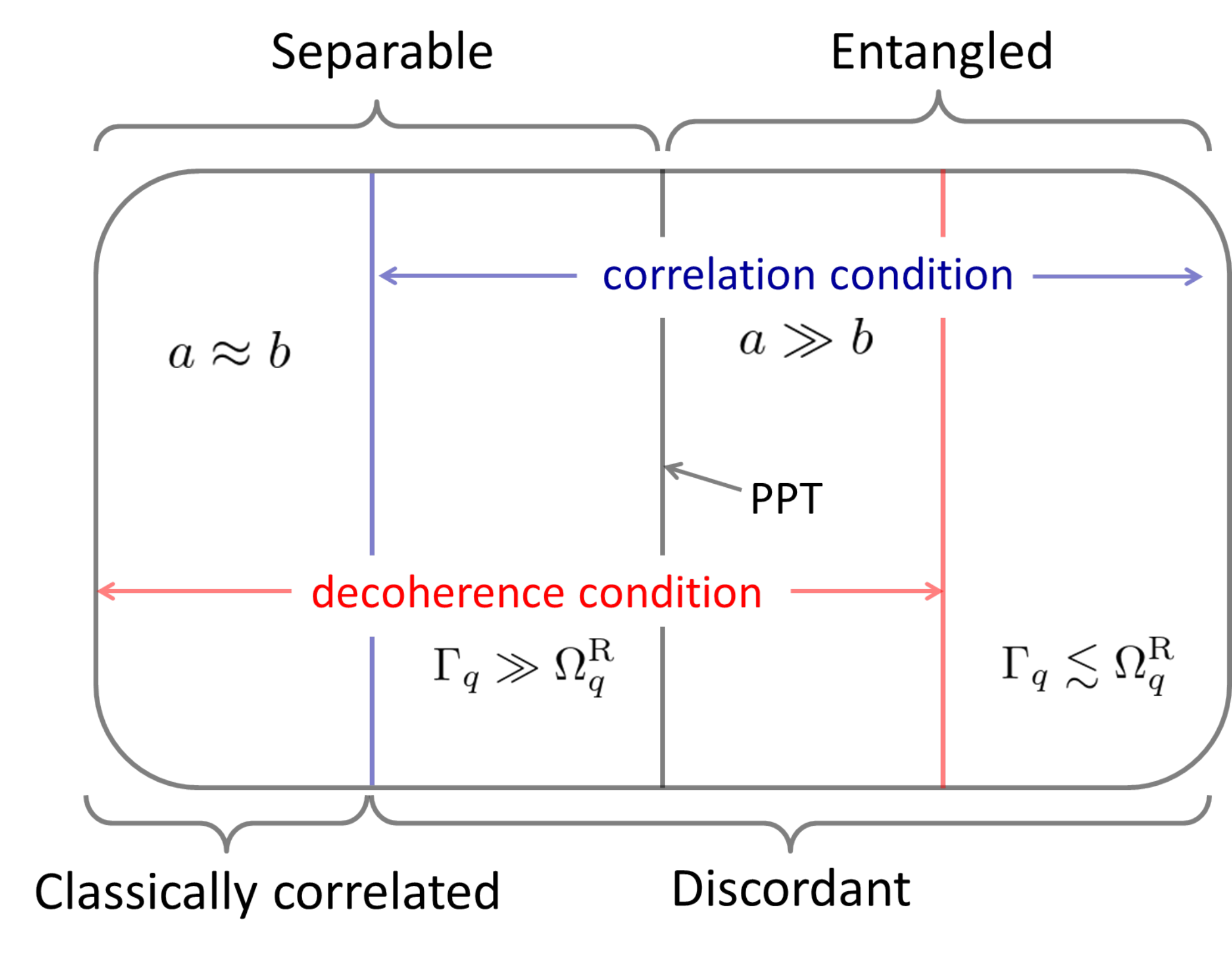}
   \caption{The relation among the quantum correlations of
     PGWs, the decoherence condition and the correlation
       condition. In the left side region of the red vertical line,
     the decoherence condition is satisfied. In the right side region
     of the blue vertical line, the correlation condition is
     satisfied.}
    \label{fig:relation}
\end{figure}

 In Fig. \ref{fig:relation}, we summarize the relation among the
 entanglement, the quantum discord of PGWs, the
 decoherence condition and the correlation condition for
 super-horizon modes. As we have mentioned after Eq.~ 
\eqref{eq:PPT2}, the oscillation of the
 angular-power spectrum implies the quantum discord of PGWs
   but does not guarantees the existence of entanglement. 
\noindent
 For the decohered state, we can
 choose the parameter $\Gamma_q(\eta_c)$ both satisfying the PPT criterion and the
 correlation condition. Thus it is also confirmed that the entanglement of PGWs is not required to
 obtain the oscillatory behavior of the angular-power spectrum
   of CMB fluctuations.

%
\section{Summary}
\label{sec:summary}

Focusing on quantum correlations, we examined the oscillation of the
angular-power spectrum of CMB fluctuations induced by PGWs. This oscillatory feature is different from the observed acoustic oscillation. The dominant contribution of the acoustic oscillation is due to primordial density perturbations not PGWs. However the oscillation caused by PGWs is related to the quantum discord of PGWs. We 
demonstrate that the constructed classical state of PGWs without
quantum discord has no oscillatory feature for the angular-power
spectrum of the CMB temperature fluctuations. For PGWs with quantum origin, 
the oscillation of the CMB power spectrum can be interpreted as the signature of the quantum discord of the PGWs.

We also investigated the
decoherence effect for super-horizon modes on the squeezing property
of PGWs. In particular, we discussed the decoherence condition and the
correlation condition \cite{Kiefer2007} in terms of quantum
correlations.  Through the comparison of the PPT criterion and the
decoherence condition, we found that the decoherece condition is not
sufficient for the separability of the PGWs state in the
matter-dominated era. Also we showed that the correlation condition
implies the quantum discord of PGWs in the matter-dominated
era. This argument is obvious because the correlation condition
ensures the squeezed Wigner function if there is no decoherence after
the horizon crossing. What we have done here is to furnish the meaning
of the correlation condition in terms of quantum discord. We expect
that the oscillatory feature of PGWs gives a hint for the question
whether PGWs are quantum or not in our observable universe.

%
\begin{acknowledgments}
Y.N. was supported in part by the JSPS KAKENHI Grant Number 19K03866.
\end{acknowledgments}
%
\appendix
\section{Proof of the equation \eqref{eq:prodq-q}}
To show that the state satisfying the assumptions 2, 3, and
  4 in the section \ref{sec:discord} requires 
  $d_{q}=0$, we consider a two-mode continuous
  variable state defined by two annihilation
  operators $\hat{a}_\text{A}$ and $\hat{a}_\text{B}$. Then we can
prove the following lemma: a two-mode classically correlated Gaussian
state $\rho_\text{AB}$ satisfies $
\text{Tr}[\hat{a}_\text{A} \hat{a}_\text{B}
\rho_\text{AB}]=\alpha_\text{A} \alpha_\text{B}$ where the parameters $\alpha_\text{A,B}$ are 
displacement of each system.
By the assumption of classically correlated (Eq.~\eqref{eq:NDstate}),
the state $\rho_\text{AB}$ is represented as
$\rho_\text{AB}=\sum_{i,j}p_{ij} \ket{\psi^{i}_\text{A}}
\bra{\psi^{i}_\text{A}} \otimes \ket{\phi^{j}_\text{B}}
\bra{\phi^{j}_\text{B}}$. Tracing out the system B, we have
$\rho_\text{A}=\sum_{i}p_{i} \ket{\psi^{i}_\text{A}}
\bra{\psi^{i}_\text{A}}$
where $p_i=\sum_{j}p_{ij}$. Since the state $\rho_\text{AB}$ is Gaussian, $\rho_\text{A}$ is a Gaussian state with the
displacement $\alpha_\text{A}$. By the orthonormal property of
$\ket{\psi^{i}_\text{A}}$, the vectors $\ket{\psi^{i}_\text{A}}$ are
eigenvectors of the state $\rho_\text{A}$. From the Williamson theorem
\cite{Williamson1936}, we can identify the state vector
$\ket{\psi^{i}_\text{A}}$ with a state
$\hat{D}_\text{A}(\alpha) \ket{N_\text{A}}$. Here
$\hat{D}_\text{A}(\alpha)$ is the displacement operator of the system
A. The parameter $\alpha_\text{A}$ does not depend on the label
  $i$, and $\ket{N_\text{A}}$ is an N-particle state defined by an
  annihilation operator $\hat{b}_\text{A}$, whose label $N$
  corresponds to the label $i$ up to the ordering. Further, the
Williamson theorem implies that there is the unitary operator
generated by the symplectic transformation such that
$\hat{a}_\text{A}= \xi_\text{A} \hat{b}_\text{A}+\eta_\text{A}
\hat{b}^{\dagger}_\text{A}$
where $\xi_\text{A}$ and $\eta_\text{A}$ are the parameters of the
symplectic transformation. The above statement holds for the
  system B. Hence we find the following equation
\begin{align}
\text{Tr}[\hat{a}_\text{A} \hat{a}_\text{B} \rho_\text{AB}] 
&=\sum_{i,j}p_{ij} \bra{\psi^{i}_\text{A}}\hat{a}_\text{A}
  \ket{\psi^{i}_\text{A}} \bra{\phi^{j}_\text{B}}\hat{a}_\text{B}
  \ket{\phi^{j}_\text{B}} \nonumber \\
&=\sum^{\infty}_{N,M=0} p_{NM}
  \bra{N_\text{A}}(\hat{a}_\text{A}+\alpha_\text{A}) \ket{N_\text{A}}
  \bra{M_\text{B}}(\hat{a}_\text{B}+\alpha_\text{B}) \ket{M_\text{B}}
%
=\alpha_\text{A} \alpha_\text{B}, \label{eq:aAaB1}
\end{align}
where we identified $\ket{\phi^{j}_\text{B}}$ with
$\hat D_\text{B}(\alpha)\ket{M_\text{B}}$. $\hat{D}_\text{B}(\alpha)$ is
the displacement operator for the system B and $\ket{M_\text{B}}$ is
an M-particle state of the system B. The equation
\eqref{eq:aAaB1} implies that the Gaussian state is a product state.

\section{Derivation of the inequality \eqref{eq:PPT2}}
We consider a two-mode Gaussian state $\rho_\text{AB}$, whose modes
are defined by the annihilation operators $\hat{a}_\text{A}$ and
$\hat{a}_\text{B}$. We introduce the vector $
\hat{\alpha}=\left[ \hat{a}_\text{A}, \hat{a}^{\dagger}_\text{A}, \hat{a}_\text{B}, \hat{a}^{\dagger}_\text{B} \right]^\text{T}$.
The covariance matrix of the state $\rho_\text{AB}$ is defined by the
Hermitian matrix $C_{ij} =\frac{1}{2} \text{Tr} [ \{
\hat{\alpha}^\dagger_{i},  \hat{\alpha}_{j} \}  \, \rho_\text{AB}]$ where $\{ \cdot , \cdot \}$ is the anti-commutator. The explicit form of the matrix $C$ is 
\begin{equation}
C=\begin{bmatrix}
\frac{1}{2} \langle \{ \hat{a}^\dagger_\text{A}, \hat{a}_\text{A} \} \rangle & \langle (\hat{a}^\dagger_\text{A})^{2}  \rangle  & \langle \hat{a}^\dagger_\text{A} \hat{a}_\text{B}  \rangle & \langle \hat{a}^\dagger_\text{A} \hat{a}^{\dagger}_\text{B}  \rangle 
\\
  & \frac{1}{2} \langle \{ \hat{a}_\text{A}, \hat{a}^\dagger_\text{A} \} \rangle  & \langle \hat{a}_\text{A} \hat{a}_\text{B}  \rangle & \langle \hat{a}_\text{A} \hat{a}^{\dagger}_\text{B}  \rangle 
\\
  &  & \frac{1}{2} \langle \{ \hat{a}^\dagger_\text{B}, \hat{a}_\text{B} \} \rangle & \langle (\hat{a}^\dagger_\text{B})^{2}  \rangle
\\
  &  &  &  \frac{1}{2} \langle \{ \hat{a}_\text{B}, \hat{a}^{\dagger}_\text{B} \} \rangle 
   \end{bmatrix},
\end{equation}
where $\langle \cdot \rangle = \text{Tr}[\cdot \, \rho_\text{AB} ]$ and the omitted components are determined by the Hermiticity. The covariance matrix satisfies the following uncertainty relation: for any $z=[z_1, z_2, z_3, z_4]^\text{T}, z_{i} \in \mathbb{C}$, 
\begin{equation}
z^\dagger C z= \frac{1}{2} \text{Tr} [ \{ (z \cdot \hat{\alpha})^\dagger,  z \cdot \hat{\alpha} \}  \, \rho_\text{AB}] = \text{Tr} [ (z \cdot \hat{\alpha})^\dagger z \cdot \hat{\alpha} \rho_\text{AB}]+\frac{1}{2} z^\dagger \Omega z  \geq \frac{1}{2} z^\dagger \Omega z, \nonumber 
\end{equation}
that is $C \geq \frac{1}{2} \Omega$ where the matrix $\Omega$ is given
by $[\hat{\alpha}_{j}, \hat{\alpha}^\dagger_{k}]= \Omega_{jk}$. The
partial transpose operation for the subsystem B is represented by
$\hat{b}_\text{A} \rightarrow \hat{b}_\text{A}$ and
$\hat{b}_\text{B} \rightarrow \hat{b}^{\dagger}_\text{B}$ \cite{Simon2000}. We denote
the partial transposed matrix as $\widetilde{C}$. Then the inequality
for the PPT criterion is $\widetilde{C} \geq \frac{1}{2} \Omega$.
The state of interest has only the two expectation values
\begin{equation}
\langle \hat{a}^\dagger_\text{A} \hat{a}_\text{A} \rangle=\langle \hat{a}^\dagger_\text{B} \hat{a}_\text{B} \rangle=n, 
\quad \langle \hat{a}_\text{A} \hat{a}_\text{B} \rangle=c.\label{eq:nc}
\end{equation}
Then the covariance matrix $C$ and its partial transposed matrix $\widetilde{C}$ are computed as 
\begin{equation}
C=\begin{bmatrix}
n+\frac{1}{2}  & 0  & 0 & c^{*} \\
0  & n+\frac{1}{2}  & c & 0 \\
0  &  c^{*} & n+\frac{1}{2} & 0 \\
c & 0 & 0 &  n+\frac{1}{2}  
   \end{bmatrix}, \quad
\widetilde{C}=\begin{bmatrix}
n+\frac{1}{2}  & 0  & c^{*} & 0 \\
0  & n+\frac{1}{2}  & 0 & c^{*} \\
c  &  0 & n+\frac{1}{2} & 0 \\
0 & c & 0 &  n+\frac{1}{2}  
   \end{bmatrix}. \label{eq:CandtildeC}
\end{equation}
From this formula of $\widetilde{C}$, we easily get the PPT criterion as $n \geq |c|$. 

\section{Derivation of the equations \eqref{eq:ycyc}, \eqref{eq:ycpic} and \eqref{eq:picpic}}
We compute the two-point functions of the decohered state \eqref{eq:decstate}.  For convenience, we use the Schr\"{o}dinger picture to calculate them: 
\begin{align}
\bra{\Psi}\hat{\Omega }^{\dagger}(\eta_\text{c},-\infty) \hat{y}^\text{I}_\lambda (\bm{q},\eta_\text{c}) \hat{y}^\text{I}_{\lambda'}  (\bm{q}',\eta_\text{c}) \hat{\Omega}(\eta_\text{c},-\infty) \ket{\Psi} 
&=\bra{\Psi}\hat{U}^\dagger (\eta_\text{c},-\infty) \hat{y}_\lambda (\bm{q}) \hat{y}_{\lambda'} (\bm{q}') \hat{U} (\eta_\text{c},-\infty) \ket{\Psi}, \label{eq:ycycS} \\
\bra{\Psi}\hat{\Omega }^{\dagger}(\eta_\text{c},-\infty) \hat{y}^\text{I}_\lambda (\bm{q},\eta_\text{c}) \hat{\pi}^\text{I}_{\lambda'}  (\bm{q}',\eta_\text{c}) \hat{\Omega}(\eta_\text{c},-\infty) \ket{\Psi} 
&=\bra{\Psi}\hat{U}^\dagger (\eta_\text{c},-\infty) \hat{y}_\lambda (\bm{q}) \hat{\pi}_{\lambda'} (\bm{q}') \hat{U} (\eta_\text{c},-\infty) \ket{\Psi}, \label{eq:ycpicS} \\
\bra{\Psi}\hat{\Omega}^{\dagger}(\eta_\text{c},-\infty) \hat{\pi}^\text{I}(\bm{q},\eta_\text{c}) \hat{\pi}^\text{I}(\bm{q}',\eta_\text{c}) \hat{\Omega}(\eta_\text{c},-\infty) \ket{\Psi} 
&=\bra{\Psi}\hat{U}^\dagger (\eta_\text{c},-\infty) \hat{\pi}_\lambda (\bm{q}) \hat{\pi}_{\lambda'} (\bm{q}') \hat{U} (\eta_\text{c},-\infty) \ket{\Psi}, \label{eq:picpicS}
\end{align}
where
$\hat{y}_\lambda (\bm{q}):=\hat{y}^\text{I}_\lambda (\bm{q},-\infty)$
and
$\hat{\pi}_\lambda (\bm{q}):= \hat{\pi}^\text{I}_\lambda
(\bm{q},-\infty)$
are the field operators and its conjugate momentum in the
Schr\"{o}dinger picture and $\hat{U}$ is the evolution operator given
by \eqref{eq:U}. The correlation function  $\langle yy \rangle$ at the
time $\eta_\text{c}$ is
\begin{align}
&\bra{\Psi}\hat{U}^\dagger (\eta_\text{c},-\infty) \hat{y}_\lambda (\bm{q}) \hat{y}_{\lambda'} (\bm{q}') \hat{U} (\eta_\text{c},-\infty) \ket{\Psi} 
= \int_y \, y_\lambda (\bm{q}) y_{\lambda'}(\bm{q}') \rho_{\eta_\text{c}}[y,y] \nonumber \\
&\qquad\qquad
 = \int_y \, y_\lambda (\bm{q}) y_{\lambda'}(\bm{q}') |\Psi^\text{BD}_{\eta_\text{c}}[y]|^{2} 
= \bra{0^\text{BD}_y} \hat{y}^\text{I}_\lambda (\bm{q},\eta_\text{c}) \hat{y}^\text{I}_{\lambda'} (\bm{q}',\eta_\text{c}) \ket{0^\text{BD}_y}. \label{eq:ycycS2} 
\end{align}
Similarly the other correlation functions $\langle y\pi \rangle$ and $\langle \pi \pi \rangle$ are computed as
\begin{align}
&\bra{\Psi}\hat{U}^\dagger (\eta_\text{c},-\infty) \hat{y}_\lambda (\bm{q}) \hat{\pi}_{\lambda'} (\bm{q}') \hat{U} (\eta_\text{c},-\infty) \ket{\Psi} 
= \int_y \, y_\lambda (\bm{q}) \Bigl[-i \frac{\delta}{\delta \tilde{y}_{\lambda'}(-\bm{q}')} \Bigr]  \rho_{\eta_\text{c}}[\tilde{y},y] \Bigl|_{\tilde{y}=y} \nonumber \\
&\qquad\qquad= \int_y  \,  y(\bm{q})\Bigl[-i \frac{\delta \Psi^\text{BD}_{\eta_\text{c}}[y]}{\delta y_{\lambda'}(-\bm{q}')} \Bigr]  \Psi^{\text{BD}*}_{\eta_\text{c}}[y] 
= \bra{0^\text{BD}_y} \hat{y}^\text{I}_\lambda (\bm{q},\eta_\text{c}) \hat{\pi}^\text{I}(\bm{q}',\eta_\text{c}) \ket{0^\text{BD}_y}, \label{eq:ycpicS2} 
\end{align}
and
\begin{align}
&\bra{\Psi}\hat{U}^\dagger (\eta_\text{c},-\infty) \hat{\pi}_\lambda (\bm{q}) \hat{\pi}_{\lambda'} (\bm{q}') \hat{U} (\eta_\text{c},-\infty) \ket{\Psi} 
= \int_y \Bigl[i \frac{\delta}{\delta y_\lambda (-\bm{q})} \Bigr] \Bigl[-i \frac{\delta}{\delta \tilde{y}_{\lambda'} (-\bm{q}')} \Bigr]  \rho_{\eta_\text{c}}[\tilde{y},y] \Bigl|_{\tilde{y}=y} \nonumber \\
&\qquad\qquad = \int_y  \, \Bigl[-i \frac{\delta \Psi^\text{BD}_{\eta_\text{c}}[y]}{\delta y_{\lambda'} (-\bm{q}')} \Bigr]  \Bigl[i \frac{\delta \Psi^{\text{BD}*}_{\eta_\text{c}}[y]}{\delta y_\lambda (-\bm{q})} \Bigr]+\Gamma_q (\eta_\text{c}) \delta_{\lambda \lambda'} \delta^{3}(\bm{q}+\bm{q}')  
\nonumber \\
&\qquad\qquad
= \bra{0^\text{BD}_y} \hat{\pi}^\text{I}_\lambda (\bm{q},\eta_\text{c}) \hat{\pi}^\text{I}_{\lambda'} (\bm{q}',\eta_\text{c}) \ket{0^\text{BD}_y} +\Gamma_q (\eta_\text{c})\delta_{\lambda \lambda'} \delta^{3}(\bm{q}+\bm{q}'), \label{eq:picpicS2} 
\end{align}
where we used the functional representation of the conjugate momentum $\hat{\pi}_\lambda (\bm{q})=-i\delta/\delta y_\lambda (-\bm{q})$. 

%

\end{document}